\documentstyle[12pt]{article}

\newcommand{\be}{\begin{equation}}
\newcommand{\ee}{\end{equation}}
\def\aprle{\buildrel < \over {_{\sim}}}
\def\aprge{\buildrel > \over {_{\sim}}}
\begin{document}

\topmargin 0pt
\oddsidemargin=-0.4truecm
\evensidemargin=-0.4truecm

\renewcommand{\thefootnote}{\fnsymbol{footnote}}

\newpage
\newpage
\vspace*{-2.0cm}
\vspace*{0.5cm}
\begin{center}
{\Large \bf Large and infinite extra dimensions}
\vspace{0.5cm}

{\large V. A. Rubakov
\footnote{E-mail: rubakov@ms2.inr.ac.ru}\\ 
}
\vspace*{0.4cm}
\em{
Institute for Nuclear Research of the Russian Academy
of Sciences, 60th October Anniversary
Prospect, 7a, Moscow, 117312, Russia
}
\end{center}
\vglue 0.8truecm
\begin{abstract}
The emphasis in the developmet of theories with more than
three spatial dimensions has recently
shifted towards ``brane world''
picture, which assumes that ordinary matter (with possible exceptions
of gravitons and other, hypothetic, particles which interact very
weakly with matter) is trapped to a three-dimensional submanifold ---
brane --- embedded in fundamental multi-dimensional space. In the
brane world scenario, extra dimensions may be large, and even infinite;
they may have effects, directly observable in current or fothcoming 
experiments.   
On the basis of simple field-theoretic models, various ideas in this 
direction are exposed at a non-expert level.

\end{abstract}
\vspace{1.cm}
\tableofcontents
\setcounter{page}{0}

\renewcommand{\thefootnote}{\arabic{footnote}}
\setcounter{footnote}{0}
\newpage
\section{Introduction}

The possibility that our space has more than three spatial dimensions
has been attracting continuing interest for many years. Strong
motivation for considering space as multi-dimensional comes from
theories which incorporate gravity in a reliable manner --- string
theory and M-theory: almost all their versions are naturally and/or
consistently formulated in space-time of more than four dimensions. In
parallel to developments in the fundamental theory, studies along more
phenomenological lines have recently lead to new insights on whether
and how extra dimensions may manifest themselves, and whether and how
they may help to solve long-standing problems of particle theory
(hierarchy problem, cosmological constant problem, etc.). These
phenomenological studies are often based on simplified (even
over-simplified) field-theoretic models, and this approach has its
advantages and disadvantages. An advantage is that, by considering 
various models, one reveals a whole spectrum of possibilities. A
disadvantage is that some (most?) of these models may have nothing to
do with fundamental theory, so it is  unclear which of these
possibilities have a chance to be realized in nature. Furthermore,
quantitative estimates are often at best order-of-magnitude, and 
in many cases are not available at all, as most of these models have
free parameters. Still, the phenomenological approach, which is the
subject of this mini-review, helps to understand how to search for
extra dimensions, if there exist ones.

An important issue in multi-dimensional theories is the mechnaism
by which  extra dimensions are hidden, so that the space-time is
effectively four-dimensional insofar as known physics is
concerned. Until recently, the main emphasis was put on theories of
Kaluza--Klein type, where extra dimensions are compact and essentially
homogeneous. 
In this picture,  it is the compactness of extra dimensions that
ensures that space-time is effectively four-dimensional at distances
exceeding the compactification scale (size of
extra-dimensions). Hence, the size of extra dimensions must be
microscopic; 
a ``common wisdom'' was that this size was roughly of
the order of the Planck scale (although compactifications at the
electroweak scale were also considered, see. e.g.,  
\cite{Volobuev:1986wc,Antoniadis:1990ew,Lykken:1996fj}). 
With the Planck length $l_{Pl} \sim 10^{-33}$~cm
and the corresponding energy scale $M_{Pl} \sim 10^{19}$~GeV,
probing extra dimensions directly appeared hopeless.

Recently, however, the emphasis has shifted towards ``brane world''
picture which assumes that ordinary matter (with possible exceptions
of gravitons and other, hypothetic, particles which interact very
weakly with matter) is trapped to a three-dimensional submanifold ---
brane --- embedded in fundamental multi-dimensional space. In the
brane world scenario, extra dimensions may be large, and even infinite;
we shall see that they may then have experimentally observable
effects.

Certainly, the potential detectability of large and infinite extra
dimensions is one of the reasons of why they are interesting. Another
reason is that lower-dimensional manifolds, $p$-branes, are inherent
in string/M-theory. Some kinds of $p$-branes are capable of carrying
matter fields; for example, $D$-branes have gauge fields residing on
them
(for a review, see Ref.~\cite{Polchinski:1996na}). 
Hence, the general idea of brane world appears naturally in M-theory
context, and, indeed, realistic brane-world models based on M-theory
have been proposed~\cite{Horava:1996qa,Lukas:1999yy}. 
Even though the phenomenological
models to be discussed in this mini-review
may have nothing to do with M-theory $p$-branes, one hopes that
some of their properties will have counterparts in the
fundamental theory. We note in this regard that the term ``brane'' has
quite different meaning in different contexts; we shall use this term
for any three-dimensional submanifold to which ordinary matter is trapped,
irrespectively of the trapping mechanism. 

The purpose of this mini-review is to expose, on the basis of simple
models, some ideas and results related to large and infinite extra
dimensions. We do not attempt at a comprehensive discussion 
of such constructions; our choice of topics will thus be very
personal, and the list of references very incomplete. Neither are we
going to present historical overview; a view on history of the brane
world scenario is presented, e.g., in Ref.~\cite{Visser:1985qm}.

\section{Kaluza--Klein picture}

To begin with,
let us outline the basic idea of the Kaluza--Klein scenario; this will
serve as a point of reference for further discussions. The simplest
case is one extra spatial dimension $z$, so that the complete set of
coordinates in (4+1)-dimensional space-time is $(x^{\mu}, z)$, 
$\mu = 0,1,2,3$. The low energy physics will be effectively
four-dimensional if the coordinate $z$ is compact with a certain
compactification radius $R$. This means that $z$ runs from $0$ to
$2\pi R$, and points $z=0$ and $z=2\pi R$ are identified. In other
words, the four-dimensional space is a cylider whose three dimensions
$x^1, x^2, x^3$  are infinite, and the fourth dimension $z$ is a
circle of radius $R$. Assuming that this cylinder is homogeneous, and
that the metric is flat, one writes a complete set of wave functions
of a free massless particle on this cylinder (e.g., solutions to
five-dimensional Klein--Gordon equation),
\[
\phi_{{\bf p}, n} = \mbox{e}^{ip_{\mu}x^{\mu}} \mbox{e}^{inz/R}\;, 
\quad n=0,\pm 1,\pm 2, \dots \;.
\]
Here $p_{\mu}$ is the (3+1)-dimensional momentum and $n$ is the
eigenvalue of (one-dimensional) angular momentum. 
Since $\phi(x,z)$ obeys $\Box_{(5)} \phi = 0$,
these quantities are
related,
\be
   p_{\mu} p^{\mu} - \frac{n^2}{R^2} = 0
\label{9*}
\ee
Hence, inhomogeneous modes with $n\neq 0$ carry energy of order $1/R$,
and they cannot be excited in low energy processes. Below the energy
scale $1/R$, only homogeneous modes with $n=0$ are relevant, and low
energy
physics is effectively four-dimensional.

From (3+1)-dimensional point of view, each Kaluza--Klein (KK)
mode
 can be interpreted as a separate type of particle with mass 
$m_n = |n|/R$, according to eq.~(\ref{9*}). Every multi-dimensional
field corresponds to a Kaluza--Klein tower of four-dimensional
particles with
increasing masses. At low energies, only massless (on the scale $1/R$)
particles can be produced, whereas at $E \sim 1/R$ extra dimensions
will show up. Since the KK partners of ordinary particles (electrons,
photons, etc.) have not been observed, the energy scale $1/R$ must be
at least in a few hundred GeV range, so in the Kaluza--Klein scenario,
the size of extra dimensions must be microscopic ($R \aprle
10^{-17}$~cm).
These properties are inherent in all models of Kaluza--Klein type
(with larger number of extra dimensions, compactifications on
non-trivial manifolds or orbifolds instead of a circle, etc.).

\section{Localized matter}
To see that it is indeed conceivable that ordinary matter may be
trapped to a brane, we present in this section simple field-theoretic
models exhibiting this property. Throughout this section we neglect
gravity; new possibilities emerging when gravitational interactions
are included will be considered later on.

\subsection{Localized  fermions}

It is fairly straightforward to construct field-theoretic models with
localized fermions. The simplest model of this sort has one extra
dimension $z$, with the brane being a domain wall
\cite{Rubakov:1983bb}
(see also Ref.~\cite{Akama:1982jy}).
Namely, let us consider a theory of one real scalar field
$\varphi$ whose action is
\be
S_{\varphi} = \int~d^4x~dz~\left[ \frac{1}{2} (\partial_A \varphi)^2
- V(\varphi) \right]
\label{r4*}
\ee
Here the subscript $A$ denotes all five coordinates, and the scalar
potential $V(\varphi)$ has a double-well shape with two degenerate
minima at
$\varphi = \pm v$,
as shown in Fig. 1.

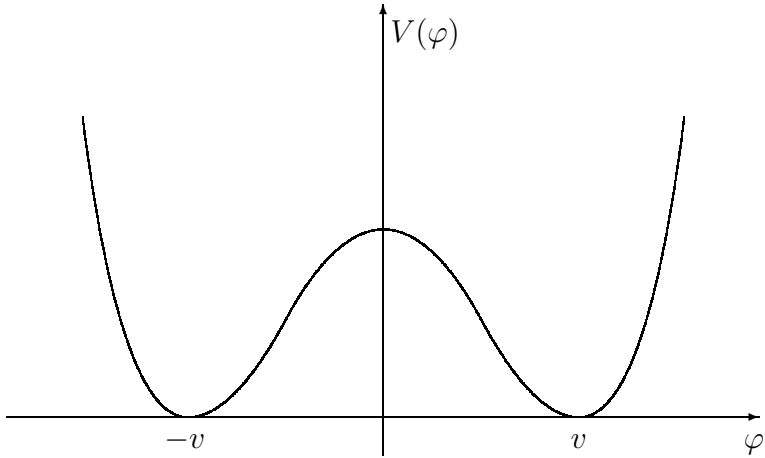
\begin{figure}[!hb]
\begin{center}
\unitlength1mm
\begin{picture}(100,60)(-50,-5)
\put(-50,0){\vector(1,0){100}}
\put(0,-5){\vector(0,1){60}}
\put(-29,-4){$-v$}
\put(25,-4){$v$}
\put(1,50){$V(\varphi)$}
\put(48,-4){$\varphi$}
\qbezier[600](-40,40)(-35,0)(-26,0)
\qbezier(-26,0)(-20,0)(-13,13)
\qbezier(-13,13)(0,37)(13,13)
\qbezier(13,13)(20,0)(26,0)
\qbezier[600](26,0)(35,0)(40,40)
\end{picture}
\caption{\label{1}Scalar potential in the model (\ref{r4*}).}
\end{center}
\end{figure}

There exists a classical solution $\varphi_c (z)$, kink, depending
on one coordinate only. This solution is sketched in Fig. 2. It has asymptotics
\[
     \varphi_c (z \to + \infty) = + v
\]
\[
     \varphi_c (z \to - \infty) = - v
\]
and describes a domain wall separating two classical vacua of the
model. Obviously, the field $\varphi_c (z)$ breaks translational
invariance along the extra dimension, but leaves the four-dimensional
Poincar\'e invariance intact. 

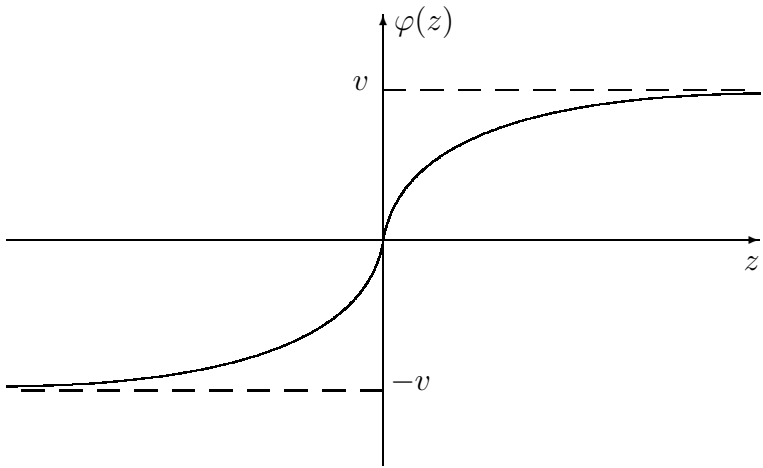
\begin{figure}[!hb]
\begin{center}
\unitlength1mm
\begin{picture}(100,60)(-50,-30)
\put(-50,0){\vector(1,0){100}}
\put(0,-30){\vector(0,1){60}}
\put(1,-20){$-v$}
\put(-4,20){$v$}
\put(1.5,28){$\varphi(z)$}
\put(48,-4){$z$}
\qbezier(-50,-19.5)(-3,-19)(0,0)
\multiput(0,-20)(-5,0){10}{\line(-1,0){3}}
\qbezier(50,19.5)(3,19)(0,0)
\multiput(0,20)(5,0){10}{\line(1,0){3}}
\end{picture}
\caption{\label{2}Domain wall solution.}
\end{center}
\end{figure}

Let us now introduce fermions into this model. We recall that fermions
in five-dimensional space-time are four-component columns, and that
the five-dimensional gamma-matrices can be chosen as follows,
\[
  \Gamma^{\mu} = \gamma^{\mu}\;, \quad \mu = 0,1,2,3 \;,
\]
\[
  \Gamma^{z} = -i \gamma^5
\]
where $\gamma^{\mu}$ and $\gamma^5$ are the standard Dirac matrices of
four-dimensional theory. Introducing the Yukawa interaction of
fermions with the scalar field $\varphi$, we write the
five-dimensional action for fermions,
\be
  S_{\Psi} = \int~d^4 x~dz~ \left( i \bar{\Psi} \Gamma^A \partial_A
  \Psi
- h \varphi \bar{\Psi} \Psi \right)
\label{13*}
\ee
Note that in each of the scalar field vacua, $\varphi = \pm v$,
five-dimensional fermions acquire a mass
\[
           m_5 = hv
\]
Let us consider fermions in the domain wall background. The
corresponding Dirac equation is
\be
 i \Gamma^A \partial_A \Psi - h \varphi_c (z) \Psi  = 0
\label{14*}
\ee
Due to unbroken four-dimensional Poincar\'e invariance, the fermion wave
functions may be characterized by four-momentum $p_{\mu}$, and we are
interested in the spectrum of four-dimensional masses $m^2 =
p_{\mu}p^{\mu}$. A key point is that there exists a zero mode
\cite{Jackiw:1976fn}, a solution to eq.~(\ref{14*})
 with $m=0$. For this mode one has 
$\gamma^{\mu} p_{\mu} \Psi_0 = 0$, and the Dirac equation (\ref{14*})
becomes
\[
\gamma^5 \partial_z \Psi_0 = h \varphi_c (z) \Psi_0
\]
The zero mode is left-handed from the four-dimensional point of view, 
\[
  \gamma_5 \Psi_0 = - \Psi_0
\]
and has the form
\be
   \Psi_0 = \mbox{e}^{- \int_0^{z}~dz'~h \varphi_c (z')}
\psi_L(p)
\label{14**}
\ee
where $\psi_L (p)$ is the usual solution of the four-dimensional Weyl
equation. The zero mode (\ref{14**}) is localized near $z=0$, i.e., at
the domain wall, and at large $|z|$ it decays exponentially,
$\Psi_0 \propto \mbox{exp}(-m_5 |z|)$.

The spectrum of four-dimensional masses  
 is shown in Fig. 3. 
Besides the chiral zero mode, there may or may not exist
bound states, but in any case the masses of the latter are
proportional to $v$ and are large for large $v$. There is also a
continuum part of the spectrum starting at $m=m_5$; the continuum
states correspond to five-dimensional fermions which are not bound to
the domain wall and escape to $|z|= \infty$.

\begin{figure}[!hb]
\begin{center}
\unitlength1mm
\begin{picture}(100,60)(-30,0)
\put(-30,0){\line(1,0){70}}
\put(0,0){\vector(0,1){60}}
\put(42,0){$m=0$}
\put(-30,1){\line(1,0){70}}
\put(1.5,58){$m_{\mathrm{fermion}}$}
\put(-30,11){\line(1,0){70}}
\put(38,6){\vector(0,1){5}}
\put(38,6){\vector(0,-1){5}}
\put(40,5){$\mathrm{const}\cdot h\nu$}
\put(-30,17){\line(1,0){75}}
\put(46,16){$m_{5} = h\nu$}
\multiput(-30,18)(0,1){20}{\line(1,0){70}}
\end{picture}
\caption{\label{3}Spectrum of four-dimensional masses of fermions 
in domain wall background. The gap between zero mode (with $m=0$) and
non-zero modes is of order $hv$. Continuum starts at 
$m=m_5 \equiv hv$.}
\end{center}
\end{figure}
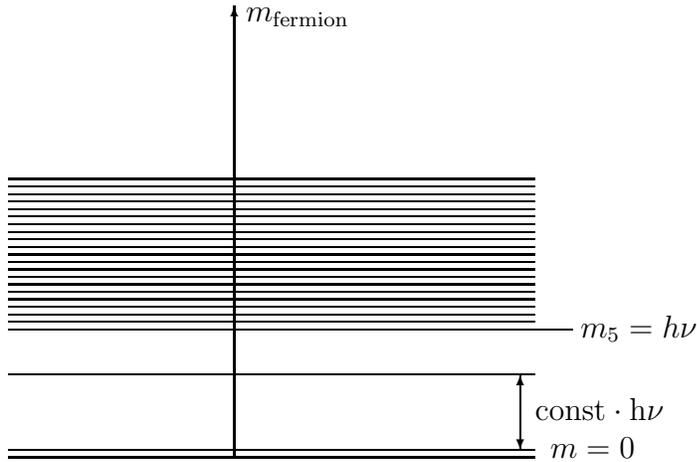

Massless four-dimensional fermions localized on the domain wall,
zero modes, are meant to mimick our matter. They propagate with the
speed of light along the domain wall, but do not move along $z$.
Of course, in realistic theories they should acquire small masses
by one or another mechanism.
At low energies, their interactions can produce only zero modes again,
so physics is effectively four-dimensional. Zero modes interacting at
 high energies, however, will produce continuum modes,
the extra dimension will open up, and
particles will be able to leave the brane, escape to $|z| = \infty$
(if the size of extra dimension 
is infinite)
and literally disappear from our world. For a four-dimensional
observer (composed of particles trapped to the brane), these high
energy processes will look like $e^+ e^- \to \mbox{nothing}$ or
$e^+ e^- \to  \gamma + \mbox{nothing}$. We shall discuss later on
whether these and similar processes are indeed possible when
gravitational and gauge interactions are taken into account, whether
they may lead to apparent non-conservation of energy, electric charge,
etc.

The above construction is straightforwardly generalized to more than
one extra dimensions. This is done by considering, instead of the
domain wall,  topological defects of higher codimensions: the
Abrikosov--Nielsen--Olesen vortex in six-dimensional space-time
($D=6$, number of extra dimensions $d=2$), 't~Hooft--Polyakov monopole
in $D=7$ ($d=3$), etc. In many cases, the existence of fermion zero
modes in the background of the topological defect is guaranteed by the
corresponding index theorem. Explicit expressions for fermion zero modes
in various backgrounds are given in 
Refs.~\cite{Jackiw:1976fn,Jackiw:1981ee,'tHooft:1976fv}. As a bonus,
the four-dimensional massless fermions localized on topological
defects are usually chiral. Furthermore, the number of fermion zero
modes may be greater than one, so from one family of multi-dimensional
fermions one can obtain several four-dimensional families. This
possibility of explaining the origin of three Standard Model
generations has been explored in
Refs.~\cite{Libanov:2001uf,Frere:2000dc} where it has been found that
reasonable pattern of masses and mixings can be obtained in a fairly
natural way.

\subsection{Localized gauge fields}

Localizing gauge fields on a brane is more difficult. The mechanism
just described does not have chance to work, at least for massless
non-Abelian fields. The reason is as follows. If the gauge field had a
zero mode whose wave function $A(z)$ is localized near the brane, the 
four-dimensional effective interaction between this localized field
and other localized fields (say, fermions) would involve overlap
integrals of the form
\be
         \int~dz~\Psi^{\dagger}_0 (z) A(z) \Psi_0 (z)
\label{18*}
\ee
where $\Psi_0$ is the fermion zero mode. We have seen that fermion
zero modes may depend on various parameters (e.g., the
coupling constant $h$ in
the example of the previous subsection: 
the width of the zero mode explicitly depends
on $h$, see eq.~(\ref{14**})). Therefore, the gauge charges in
effective four-dimensional theory would be different, at least in
principle,  for different types of particles, and they would take
arbitrary values depending on the overlap integrals like (\ref{18*}).
This is impossible in non-Abelian gauge theories where the gauge charge is
quantized, i.e., it depends only on  representation  to which
a matter field belongs, up to a factor common to all fields.

Any mechanism of the localization of (non-Abelian) massless gauge fields 
must automatically preserve charge universality, i.e., ensure that
gauge charges of all four-dimensional particles are the same (up to
group representation factors) irrespectively of the structure of
their
wave functions in transverse directions or other details of
a mechanism that binds these particles 
to our brane. To the best of
author's knowledge,  in the absence of gravity\footnote{If gravity
is turned on, other mechanisms may appear, as we discuss later.}, the
only  field-theoretic
mechanism\footnote{An alternative mechanism proposed in
Refs.~\cite{Dvali:2000hr,Dvali:2001xg,Dvali:2001rx} does not, in
general, preserve charge universality, and, therefore, it
has problems in
non-Abelian case. These will be discussed elsewhere.} of gauge field
localization which ensures charge universality,
is that of Ref.~\cite{Dvali:1997xe}. It has been proposed to consider
a gauge theory which is in confinement phase in the bulk (outside the
brane), whereas there is no confinement on the brane. Then the
electric field of a charge residing on the brane will not penetrate
into the bulk, the multi-dimensional Gauss' law will reduce to the
four-dimensional Gauss' law, and the electric field on the brane will
fall off according to the four-dimensional Coulomb law, $E \propto
1/r^2$.

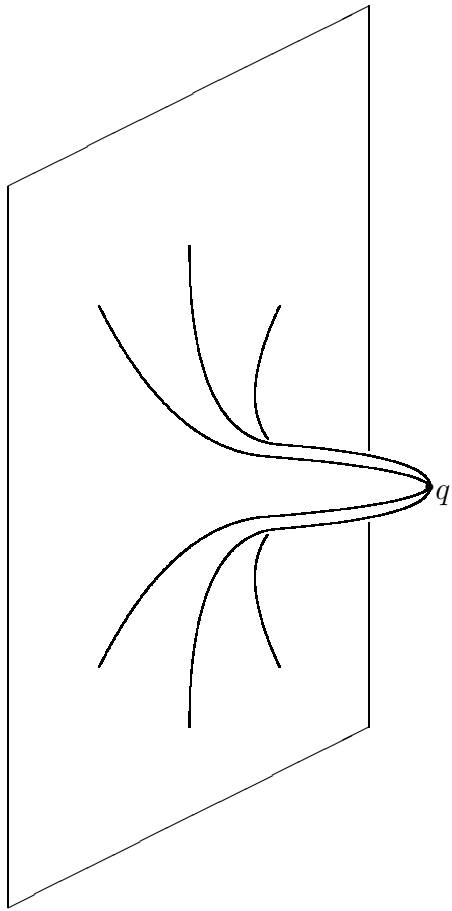
\begin{figure}[h]
\begin{center}
\unitlength.8mm
\begin{picture}(70,140)(-30,-70)
\put(-30,-70){\line(2,1){60}}
\put(-30,-70){\line(0,1){120}}
\put(-30,50){\line(2,1){60}}
\put(30,-40){\line(0,1){34}}
\put(30,80){\line(0,-1){74}}
\put(40,0){\circle*{1.5}}
\put(41,-2){$q$}
\qbezier(0,40)(0,8)(15,7)\qbezier(15,7)(40,5)(40,0)
\qbezier(0,-40)(0,-8)(15,-7)\qbezier(15,-7)(40,-5)(40,0)
\qbezier(-15,30)(-3,6)(13,5)\qbezier(13,5)(38,3)(40,0)
\qbezier(-15,-30)(-3,-6)(13,-5)\qbezier(13,-5)(38,-3)(40,0)
\qbezier(15,30)(8,15)(13,8)
\qbezier(15,-30)(8,-15)(13,-8)
\end{picture}
\caption{\label{4}Charge $q$, displaced from the brane, is connected to the brane
by a flux tube.}
\end{center}
\end{figure}

A dual analogue of this situation is an inhomogeneous superconductor
with superconductivity destroyed on a plane (i.e., Cooper pair
condensate vanishing on the plane). Magnetic monopoles placed far away
from the plane will experience confinement (there will be an Abrikosov
vortex connecting a monopole and anti-monopole), while monopoles
residing on the plane will interact according to the two-dimesnional
Coulomb law. 

Charge universality in this setting is ensured by confinement in the
bulk. If charge is displaced from the brane, a vortex connecting
this charge to the brane and carrying all of the flux will be formed
as shown in Fig. 4. 
The gauge field induced by this charge on the brane at large
distances will be independent of the position of this charge in extra
dimensions and will be identical to the three-dimensional Coulomb
field of a charge placed exactly on the brane.

This picture can be made explicit \cite{Dubrub} by considering
an Abelian model of dual superconductivity in an arbitrary number of
dimensions, along the lines of Ref.~\cite{Quevedo}. 
The corresponding calculations are rather involved, and we do not
reproduce them here.

It is worth noting that
confinement in the bulk may
imply that all states propagating in the bulk are heavy. If the
corresponding mass gap is large enough, light particles carrying gauge
charges will be bound to the brane, 
and bulk modes will not be excited at low energies. Hence, the
mechanism of Ref.~\cite{Dvali:1997xe} is simultaneously a mechanism of
trapping matter (fermions, Higgs bosons, etc.) to the brane,
alternative to that discussed in Section 3.1.

We would like to warn the reader, however, that it is not known
whether non-Abelian field theories, exhibiting confinement,  
exist at all in
more than four space-time dimensions. So, in field theory context, the
mechanism of Ref.~\cite{Dvali:1997xe} in somewhat up in the air.
On the other hand, the picture with confinement in the bulk and no
confinement on the brane has a lot of similarities with the localization
of gauge fields on  $D$-branes of string/M-theory.

\section{Large extra dimensions}

\subsection{Size of extra dimensions}

Localization of matter on a brane explains why low energy physics is
effectively four-dimensional insofar as all interactions except
gravity are concerned. To include gravity, one may proceed in
different ways. One approach
\cite{Arkani-Hamed:1998rs,Antoniadis:1998ig}, hereafter called ADD,
is to neglect the brane tension (energy density per unit three-volume
of the brane) and consider compact extra dimensions. In this way the
Kaluza--Klein picture is reintroduced. The size of extra dimensions
$R$ need not, however, be microscopic (we assume for simplicity that the
sizes of all extra dimensions are of the same order). Indeed, the
distances at which non-gravitational interactions cease to be
four-dimensional are determined by the dynamics on the brane, and may
be much smaller than $R$. Only gravity becomes multi-dimensional at
scales just below $R$. The four-dimensional law of gravitational
attraction has been established experimentally down to
distances\footnote{Until recently, the distance down to which the
Newton law was established experimentally was in the several millimeter
range \cite{Mitrofanov,Su:1994gu}, for a review see
Ref.~\cite{Long:1999dk}. The new round of experiments was stimulated
 precicely by the idea that extra dimensions may be large.}
of about
0.2~mm \cite{Hoyle:2001cv}, so the size of extra dimensions is allowed
to be as large as 0.1 mm. 

This possibility opens up a new way to address the hierarchy problem
\cite{Arkani-Hamed:1998rs,Antoniadis:1998ig}, the problem of why the
elctroweak scale (of order $M_{EW} \sim 1$~TeV) is so different from
the Planck scale ($M_{Pl} \sim 10^{16}$~TeV). In multi-dimensional
theories, the four-dimensional Planck scale is not a fundamental
parameter. Rather, the mass scale of multi-dimensional gravity, which
we denote simly by  $M$,
is fundamental, as it is this latter scale that enters the full
multi-dimensional gravitational action,
\be
  S = - \frac{1}{16\pi G_{(D)}} \int~d^D X~\sqrt{g^{(D)}} R^{(D)}
\label{25*}
\ee
where 
\[
    G_{(D)} = \frac{1}{M^{D-2}} \equiv \frac{1}{M^{d+2}}
\]
is the fundamental $D$-dimensional Newton's constant, $d=D-4$ is the
number of extra dimensions, and $d^D X = d^4 x~d^d z$.

In ADD picture, the long-distance four-dimensional gravity is mediated
by the graviton zero mode (cf. Section 2) whose wave function is
homogeneous over extra dimensions. Hence, the four-dimensional
effective action describing long-distance gravity is obtained from
eq.~(\ref{25*}) by taking the metric to be independent of extra
coorinates $z$. The integration over $z$ is then trivial, and  the
effective four-dimensional gravitational action is
\[
   S_{eff} = \frac{V_d}{16\pi G_{(D)}} \int~d^4 x~\sqrt{g^{(4)}} R^{(4)}
\]
where $V_d \sim R^d$ is the volume of extra dimensions. We see that
the four-dimensional Planck mass is, up to a numerical factor of
order one, equal to
\be
    M_{Pl} = M (MR)^{\frac{d}{2}}
\label{26*}
\ee
If the size of extra dimensions is large compared to the fundamental
length $M^{-1}$, the Planck mass is much larger than the fundamental
gravity scale $M$.

One may push this line of reasoning to extreme and suppose that the
fundamental gravity scale is of the same order as the electroweak
scale, $M \sim 1$~TeV. Then the hierarchy between $M_{Pl}$ and
$M_{EW}$ is entirely due to the large size of extra dimensions. The
hierarchy problem becomes now the problem of explaining why $R$ is
large. This is certainly an interesting reformulation.

Assuming that $M \sim 1$~TeV, one calculates from eq.~(\ref{26*})
the value of $R$,
\be
    R \sim M^{-1} \left(\frac{M_{Pl}}{M}\right)^{\frac{2}{d}}
    \sim 10^{\frac{32}{d}}\cdot 10^{-17}\;\; \mbox{cm}
\label{27*}
\ee
For one extra dimension one obtains unacceptably large value of $R$.
An interesting case is $d=2$ when roughly $R \sim 1$~mm. This
observation \cite{Arkani-Hamed:1998rs} stimulated recent activity in
experimental search for deviations from Newton's gravity law  at
sub-millimeter distances. As we shall discuss later, the mass scale as
low as $M \sim 1$~TeV is in fact excluded, for $d=2$, by astrophysics
and cosmology; a more realistic value  $M \sim 30$~TeV implies
$R \sim 1 - 10$~$\mu$m. This motivates search for deviations from
Newton's law in a micro-meter range, which is difficult but not
impossible \cite{Long:2000xa,braginsky}.

For $d>2$, eq.~(\ref{27*}) results in smaller values of $R$. For
example, for $d=3$ and  $M \sim 1$~TeV one obtains $R \sim
10^{-6}$~cm. Search for violation of Newton's law at these scales
appears hopeless. For $d=6$ (full dimensionality of space time $D=10$,
as suggested by superstring theory), one has $R \sim 10^{-12}$~cm,
which is still much larger than the elctroweak scale, 
$(1~\mbox{TeV})^{-1} \sim 10^{-17}$~cm. We note, however, that the
compactification scales of different extra dimensions are not
guaranteed to
be of the same order; if some of these are much smaller than the
others, the situation with deviations from Newton's gravity in spaces
with $d>2$ may be similar to that of $d=2$. In other words, deviations
from Newton's gravity law may occur in micro-meter 
range even for $d>2$.

\subsection{Light KK gravitons:
colliders, cosmology and astrophysics}

If the fundamental gravity scale is indeed in the TeV range, one
expects that extra dimensions should start to show up in collider
experiments at energies approaching this scale. In the picture
described in this section, extra dimensions are felt exclusively by
gravitons capable of propagating in the bulk. Hence, the most
distinctive feature of this scenario is the possibility to emit
gravitons into the bulk; this process has strong dependence on the
center-of-mass energy of particles colliding on the brane and has
large probablity at energies comparable to the fundamental gravity
scale.

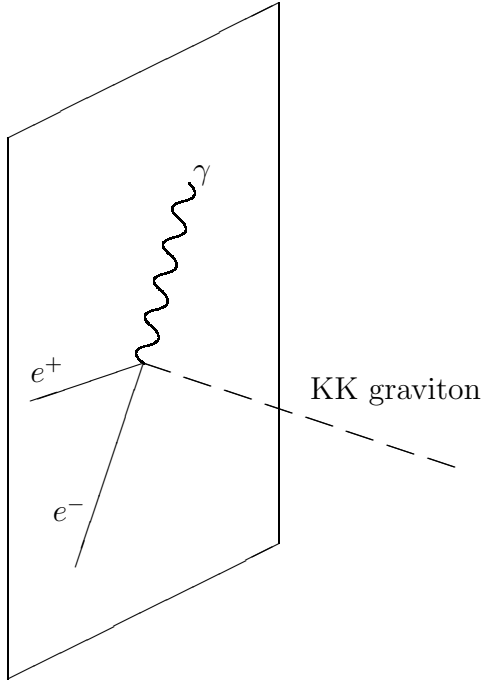
\begin{figure}[!ht]
\begin{center}
\unitlength0.6mm
\begin{picture}(70,150)(-30,-70)
\put(-30,-70){\line(2,1){60}}
\put(-30,-70){\line(0,1){120}}
\put(-30,50){\line(2,1){60}}
\put(30,-40){\line(0,1){120}}
\put(0,0){\line(-1,-3){15}}
\put(0,0){\line(-3,-1){25}}
\multiput(0,0)(9,-3){8}{\line(3,-1){6}}
\put(37,-8){KK graviton}
\put(11,41){$\gamma$}
\put(-25,-4){$e^{+}$}
\put(-20,-35){$e^{-}$}
\qbezier(0,0)(-4,3)(1,4)\qbezier(1,4)(5,5)(2,8)
\qbezier(2,8)(-2,11)(3,12)\qbezier(3,12)(7,13)(4,16)
\qbezier(4,16)(0,19)(5,20)\qbezier(5,20)(9,21)(6,24)
\qbezier(6,24)(2,27)(7,28)\qbezier(7,28)(11,29)(8,32)
\qbezier(8,32)(4,35)(9,36)\qbezier(9,36)(13,37)(10,40)
\end{picture}
\caption{\label{5}Emission of a graviton into extra dimensions (or, equivalently,
creation of a KK graviton) in the process 
$e^+ e^- \to \gamma + \mbox{graviton}$. Electron, positron and photon
propagate along the brane, the graviton escapes into the bulk.}
\end{center}
\end{figure}

From the four-dimensional viewpoint, emission of gravitons into extra
dimensions corresponds to the production of Kaluza--Klein gravitons.
One process of this type is shown in Fig. 5.
Each of KK graviton states interacts with matter on the brane with
four-dimensional gravitational strength. Indeed, the quadratic
action for each type of KK graviton, and its interaction with matter
on the brane are schematically (omitting all indices, tensor
structure, etc.) written as
\[
    S_{\bf n} = \frac{1}{16\pi G_{(D)} }
\int~d^DX~ [\partial h(x) \mbox{e}^{i {\bf q_n}{\bf z}}]^{*}
 [\partial h(x) \mbox{e}^{i {\bf q_n}{\bf z}}] 
+\int~d^4 x~h(x)T(x) 
\]
where ${\bf q_n}$  are the discrete momenta along extra dimensions
(in the case of toroidal compactification with equal sizes of extra
dimensions one has
${\bf q_n} = {\bf n}/R$, ${\bf n} = (n_1, \dots n_d)$)
and $T_{\mu \nu}$ is the energy-momentum tensor of matter on the
brane. 
The
integration over $z$ again gives the volume factor $V_{(d)}$
in front of the first term,
so the coupling of each type of KK graviton is determined by the
four-dimensional Planck mass.

Even though the coupling of every KK graviton is weak,
the total emission rate of KK gravitons is large at energies
approaching $M$ due to large number of KK graviton states. The
produced KK gravitons will fly away from a detector, so the typical
collider processes  will involve missing energy, for
example
\be
   e^+ e^- \to \gamma + \lefteqn{E}\,\raisebox{0.15ex}{/}_T
\label{30*}
\ee
or
\be
   q \bar{q} \to \mbox{jet} + \lefteqn{E}\,\raisebox{0.15ex}{/}_T
\label{30**}
\ee
The cross section of production of a KK graviton of a given type in
the process (\ref{30*}) is of order $\alpha/M_{Pl}^2$, so the total
cross section is of order
\[
   \sigma( e^+ e^- \to \gamma + 
\lefteqn{E}\,\raisebox{0.15ex}{/}_T
) \sim \frac{\alpha}{M_{Pl}^2} N(E)
\]
where $E$ is the center-of-mass energy, and $N(E)$ is the number of
spieces of KK gravitons with masses below $E$. Since the momenta along
extra dimensions, ${\bf q_ n}$, are quantized in units of $1/R$,
and the KK graviton masses are $m_n = |{\bf q_ n}|$, one has
\be
  N(E) \sim (ER)^d
\label{31*}
\ee
Making use of the relation (\ref{26*}) one obtains
\[
 \sigma( e^+ e^- \to \gamma + 
\lefteqn{E}\,\raisebox{0.15ex}{/}_T
) \sim \frac{\alpha}{E^2}
\left(\frac{E}{M}\right)^{d+2}
\]
Hence, the cross section indeed rapidly increases with the
center-of-mass energy, and at $E \sim M$ becomes comparable
to the electromagnetic cross sections.

The processes (\ref{30*}), (\ref{30**}) have been analyzed in detail
in Refs.~\cite{Giudice:1999ck,Giudice:2001av}. It has been found that
CERN LHC, as well as $e^+e^-$ collider with center-of-mass energy
1~TeV, will probe the fundamental gravity scale $M$ 
up to several TeV, the
precise number depending on $d$, the number of extra dimensions.

Another effect of extra dimensions for collider physics is the
exchange of virtual KK gravitons
\cite{Arkani-Hamed:1999nn,Giudice:1999ck,Nussinov:1999jt,Hewett:1999sn,Han:1999sg}. 
The search for this effect
at future colliders will also be sensitive to the scale $M$ in
multi-TeV range.

Light KK gravitons are essentially model-independent feature of the
ADD scenario (see, however, Ref.\cite{Kaloper:2000jb}).
If $M \sim 1$~TeV is indeed the fundamental scale
of the theory, one may expect that physics at this scale is very rich,
in particular, that there exist new particles with masses
of order $M$, either bound to the brane or propagating in the bulk.
These particles may be fairly strongly coupled to ordinary matter; in
particular, they may well carry gauge charges, and behave like heavy
electrons, quarks, vector or scalar bosons. Their properties are more
model-dependent, but in any case, current and future collider
searches for
such states are of interest from this point of view. The study of
manifestations of these heavy states has been performed, in string
theory context, in
Refs.~\cite{Kakushadze:1999wp,Shiu:1999iw,Cullen:2000ef}.

Coming back to light KK gravitons, one notes that they 
may have important effects in cosmology and
astrophysics. In the early Universe, they can be produced at high
enough
temperatures, and therefore may destroy the standard Big Bang picture
\cite{Arkani-Hamed:1999nn}. Consistency with the Big Bang
nucleosynthesis, as well as the present composition of the Universe
impose strong bounds on the maximum temperature of the Universe
\cite{Arkani-Hamed:1999nn}, as we shall now see. 

At high enough temperature, $T\gg 1/R$, the creation rate (per
unit time per unit volume) of one KK graviton spieces of mass 
$m_{\bf n} \aprle T$ is estimated as
\[
   \Gamma \sim \frac{T^6}{M_{Pl}}
\]
where the factor $M_{Pl}^{-2}$ comes from the strength of the
graviton--matter interaction, and the dependence on temperature is
restored on dimensional grounds. Taking into account the number of
KK states, cf. eq.~(\ref{31*}), one obtains the estimate of the total
rate of creation of KK gravitons,
\be
  \frac{dn}{dt} \sim \frac{T^6}{M_{Pl}} (TR)^d \sim 
    T^4 \left(\frac{T}{M}\right)^{2+d}
\label{33*}
\ee
where the latter relation comes from eq.~(\ref{26*}). Assuming that
the Universe expands in the standard way\footnote{If KK gravitons
dominated the expansion of the Universe, eq.~(\ref{33*}) would not
hold. It is straightforward to see, however, that such a scenario is
not
viable, so we do not consider this case.},
\be
    H = \frac{T^2}{M_{Pl}^{*}}
\label{33**}
\ee
with 
$M_{Pl}^{*} = M_{Pl}/(1.66 g_{*}^{1/2}) \sim (\mbox{a~few})\cdot
10^{18}$~GeV,
where $g_{*}$ is the effective number of degrees of freedom, one finds
the total number density of KK gravitons created in the Hubble time
$H^{-1}$, 
\[
   n(T) \sim T^2 M_{Pl}^{*} \left(\frac{T}{M}\right)^{2+d}
\]
Even though the creation rate (\ref{33*}) is fairly small at
$T \ll M$, the total number of gravitons may be large because of slow
expansion rate (\ref{33**}). 

A stringent bound on the maximum temperature $T_*$, that ever occurred
in the Universe after inflation, emerges if one takes the ADD picture
literally, i.e., assumes that KK gravitons survive in the bulk. Most
of the gravitons created at temperature $T_*$ have masses of order
$T_*$. Below this temperature they are non-relativistic, and their
number density scales as $T^{3}$. Hence, at the nucleosynthesis epoch
($T_{NS} \sim 1$~MeV) the mass density of KK gravitons is of order
\be
   \rho_{grav}(T_{NS}) \sim \left(\frac{T_{NS}}{T_*}\right)^3
\cdot T_* n(T_*) \sim T_{NS}^3 M_{Pl}^*
\left(\frac{T_*}{M}\right)^{2+d} 
\label{34*}
\ee
Requiring that this energy density is lower than the energy density of
one massless spieces (otherwise the standard Big Bang nucleosynthesis
would fail), i.e., that $\rho_{grav}(T_{NS}) \aprle T_{NS}^4$, one
obtains
\[
   T_* \aprle M \left(\frac{T_{NS}}{M_{Pl}^{*}}\right)^{\frac{1}{2+d}}
\sim M \cdot 10^{-\frac{21}{2+d}}
\]
For $d=2$ and $M=1$~TeV one finds that the maximum temperature should
not exceed 10~MeV, and even for $d=6$ one obtains fairly low maximum
temperature, $T_* \aprle 1$~GeV. 

Even stronger bounds on $T_*$ are obtained by requiring that the
present mass density of KK gravitons does not exceed the actual energy
density, which is close to the critical density, and that decaying KK
gravitons do not produce too much diffuse photon background 
\cite{Arkani-Hamed:1999nn}. For $d=2$, the very fact that the Universe
underwent the nucleosynthesis epoch (i.e., that $T_* \aprge
(\mbox{a~few})$~MeV) pushes the fundamental gravity scale up to 
$M \aprge  (\mbox{a~few}) \cdot 10$~TeV.

Low maximum temperature of the  Universe (say, in the range
10~MeV -- 1~GeV) does not directly contradict cosmological data:
we know for sure that the Universe underwent the standard hot Big Bang
evolution at the nucleosynthesis epoch, but have no observational
handle on higher temperature epochs\footnote{This has been explored in
detail in Ref.~\cite{Giudice:2000ex}.}. With low $T_*$, however, one
has to invoke fairly exotic mechanisms of baryogenesis and inflation,
which are possible but not very appealing.

We note in passing that we have assumed in the
above discussion that KK
gravitons do not decay before the nucleosynthesis epoch. This is
correct if nothing happens to gravitons emitted into the bulk: the
width of a graviton with mass of order $T_*$ with respect to decay
into ordinary particles (photons, $e^+e^-$-pairs, etc.) is of order
$T_*^3/M_{Pl}^2$, which, for $T_* \aprle 1$~GeV (and even for
substantially 
larger $T_*$),
is much smaller than the expansion rate before
and at nucleosynthesis, $H \sim T_{NS}^2/M_{Pl}^*$. One might invent
mechanisms of faster decay of KK gravitons into something massless in
the bulk (or on a different brane). Then the energy density of the
latter would scale as $T^{4}$, and the cosmological constraints on
$T_*$ would be weaker. Still, $T_*$ is required to be rather
low. Indeed, instead of eq.~(\ref{34*}) one would have
\[
   \rho_{extra} (T_{NS}) \sim
 \left(\frac{T_{NS}}{T_*}\right)^4
\cdot T_* n(T_*) \sim \frac{T_{NS}^4}{T_*} M_{Pl}^*
\left(\frac{T_*}{M}\right)^{2+d} 
\]
Requiring again that $\rho_{extra} (T_{NS}) \aprle T_{NS}^4$, one
obtains
\[
   T_* \aprle M \left( \frac{M}{M_{Pl}^*}\right)^{\frac{1}{2+d}}
\]
For $M \sim 1$~TeV and $d=2$ one again obtains $T_* \aprle 10$~MeV,
whereas for $d=6$ one has $T_* \aprle 10$~GeV. This model-independent
estimate shows that the maximum temperature must be rather low
irrespectively of  the fate of emitted KK gravitons.

Light KK gravitons are potentially dangerous for astrophysics as well,
as they may be produced by stars or supernovae, take away energy, and
hence contradict observational data
\cite{Arkani-Hamed:1999nn}. Strong bounds on the fundamental
scale $M$ are obtained in this way for $d=2$ only, as for larger
number of extra dimensions, the number of KK graviton states with
small masses is suppressed, cf. eq.~(\ref{31*}). As an example, by
requiring that the emission of gravitons during the collapse of
SN1987a is not the dominant cooling process (otherwise no neutrinos
would be produced, in contradiction to observations), one obtains
\cite{Arkani-Hamed:1999nn,Cullen:1999hc,Barger:1999jf,Hanhart:2001er}
\[
   M > 30~\mbox{TeV}
\]
which is comparable to the cosmological bounds. Note, however
that this bound is obtained without any assumptions concerning the
lifetime of KK gravitons. As we alredy mentioned, with so high scale
$M$, the deviations from Newton's law are only allowed well below 1~mm 
--- more realistically, at distances at most in 1 to 10 $\mu$m range.
Even stronger bound is obtained in Ref.~\cite{Hannestad:2001jv} under
the assumption that KK gravitons produced during suprenovae collapses
decay into
usual photons (and not, say, into particles residing on  other
branes).

Thus, the ADD picture predicts interesting phenomena at TeV energy
scale: its model-independent feature is the existence of light KK
gravitons which would show up at colliders either directly, in
processes (\ref{30*}), (\ref{30**}), or indirectly, through contact
interactions induced by the exchange by virtual KK
gravitons. Model-dependent features include heavy partners of ordinary
particles. Cosmology in the ADD scenario is not, however, very
appealing: the maximum temperature of the Universe must be below
10~GeV, so one has to rely upon fairly exotic mechanisms of
baryogenesis and inflation.

\subsection{Sterile neutrinos in the bulk}

Besides gravitons, there may exist other light fields not bound to the
brane and freely propagating in the bulk. The large number of their KK
states of very small masses will be phenomenologically acceptable only
if these fields are neutral under the Standard Model gauge group
$SU(3)_c \times SU(2)_L \times U(1)_Y$. An interesting candidate is a
neutral fermion which couples to conventional left-handed neutrinos
and the Standard Model Higgs field, both trapped to the brane. This
interaction induces naturally small Dirac neutrino masses
\cite{Arkani-Hamed:1998vp,Dienes:1999sb}. 

The $D$-dimensional action involving $D$-dimensional fields
$\Psi_{\nu}$ and $H$, whose zero modes describe four-dimensional flavor
neutrino and Higgs boson, respectively, and a neutral bulk fermion $\Psi$
is
\be
S_{(D)} = \int~d^4 x~d^d z~\bar{\Psi} \Gamma^A \partial_A \Psi
- \kappa  \int~d^4 x~d^d z~\bar{\Psi}_{\nu} H \Psi + \mbox{h.c.} +
\dots 
\label{nu2*}
\ee
where dots denote terms without the neutral fermion $\Psi$. The
coupling constant has dimension $(\mbox{mass})^{-d/2}$. Let us consider
the effective four-dimensional theory of the usual neutrino and Higgs
field interacting with  KK modes of the bulk fermion. For
$\Psi_{\nu}$ and $H$ we write
\begin{eqnarray}
   \Psi_{\nu} (x,z) &=& \nu (x) \psi_0 (z)
\nonumber \\
  H(x,z) &=& h(x) H_0 (z)
\end{eqnarray}
where $\nu(x)$ and $H(x)$ are four-dimensional fields and $\psi_0 (z)$
and $H_0 (z)$ are wave functions in transverse dimensions. The latter
concentrate near the brane, $z=0$, and are normalized to unity. For
the neutral bulk fermion, we have  KK decomposition 
\[
  \Psi (x,z) \propto \sum_{{\bf n}=0}^{\infty}~\psi_{\bf n} (x) 
\frac{1}{R^{\frac{d}{2}}}
  \mbox{e}^{i{\bf nz}/R} 
\]
where the factor $R^{-d/2}$ is introduced for canonical normalization
of the four-dimensional fermion $\psi_n(x)$. The effective
four-dimensional theory is then described by the effective action
\be
  S_{eff} =  \sum_{{\bf n}=0}^{\infty}~
\int~d^4 x~\left( \bar{\psi}_{\bf n} \gamma^{\mu}
  \partial_{\mu} \psi_{\bf n} - 
m_{\bf n} \bar{\psi}_{\bf n} \psi_{\bf n} 
 - \frac{\kappa^{\prime}}{R^{\frac{d}{2}}} \bar{\nu}_L h \psi_{{\bf n},R}
\right) + \dots
\label{nu3**}
\ee
where $m_{\bf n} \sim |{\bf n}|/R$ and we assume that the neutrino zero mode is
four-dimensionally left-handed (which is necessary for realistic
phenomenology and is natural in field-theoretic models, as discussed in
Section 3). The coupling $\kappa^{\prime}$ involves the overlap
integral of $\psi_0 (z)$ and $H_0 (z)$ which does not contain
$R$-dependent factors, and hence is naturally of order one, so that
\[
  \kappa^{\prime} \sim \kappa
\]
Once the Higgs field $h$ acquires vacuum expectation value 
$v \sim M_{EW}$, the last term in eq.(\ref{nu3**}) induces the
Dirac neutrino mass
\be
  m_{\nu, LR} = \frac{\kappa^{\prime} v}{R^{\frac{d}{2}}}
\label{nu3*}
\ee
For $n=0$ (the lowest KK state of the bulk fermion), this is the only
mass term in the effective theory; this implies that if
 $m_{\nu, LR} \ll 1/R$ (i.e.,
$m_{\nu, LR} \ll m_{\bf n}$ for ${\bf n} \neq 0$), then
the KK modes with ${\bf n}\neq 0$ are
irrelevant at very low energies, and the four-dimensional theory
reduces to the theory of a Dirac neutrino with mass (\ref{nu3*}), plus
other sterile fermions which decouple.

Since the Dirac mass (\ref{nu3*}) is suppressed by the size of extra
dimensions, it is naturally small. Taking, as a crude estimate, 
\[
  \kappa^{\prime} \sim \kappa \sim M^{-\frac{d}{2}}
\]
where $M$ is again the fundamental gravity scale,
one finds
\[
 m_{\nu, LR} \sim  \frac{v}{(M R)^{\frac{d}{2}}}
\]
Recalling the relation (\ref{26*}), one obtains
\[
 m_{\nu, LR} \sim  \frac{v M}{M_{Pl}}
\]
which is of order $10^{-4}$~eV for $M \sim (\mbox{a~few})\cdot$~TeV.
This estimate suggests that the neutrino masses may naturally fall in
the ballpark suggested by solar and atmospheric neutrino data.

It is worth noting that the reason for smallness of the neutrino
masses in this picture has precisely the same origin as the smallness
of the four-dimensional Newton's constant: the bulk fields spread over
the whole space of extra dimensions, and thus interact very
weakly with
matter residing on the brane.

Refinement of this picture has lead to a number of interesting effects
\cite{Dienes:1999sb,Dvali:1999cn,Mohapatra:1999zd,Barbieri:2000mg,Lukas:2000wn,Cosme:2000ib}.
The Yukawa coupling in eq.~(\ref{nu2*}) is not, in general,
flavour-diagonal. This property, and especially mixing of flavor
neutrinos $\nu_e$, $\nu_{\mu}$, $\nu_{\tau}$ 
 with higher KK states $\psi_n$ induce neutrino oscillation patterns
which are often very non-trivial. It has been found that in a
reasonably wide region of the parameter space, these patterns are
consistent with experimental data on neutrino oscillations, and that
this mechanism can be discriminated, in future experiments, against
conventional four-dimensional mechanisms. There are, however, strong
constraints on this scenario, notably, coming from SN~1987a 
\cite{Barbieri:2000mg,Lukas:2000wn}. These and other issues are
reviewed, e.g., in Ref.~\cite{Perez-Lorenzana:2000hf}.

\subsection{Unification of couplings}

A nice property of the four-dimensional Minimal Supersymmetric
Standard Model (MSSM), and many of its extensions, is that the gauge
couplings $\alpha_i$, $i=1,2,3$, corresponding to the gauge groups
$U(1)_Y$, $SU(2)_L$ and $SU(3)_c$, respectively, unify at the Grand
Unification scale $M_{GUT} \sim 10^{16}$~GeV
($\alpha_1$ is actually defined as $(5/3) \alpha_Y$). 
This occurs through the logarithmic running of these couplings
according to the renormalization group.
Gauge coupling unification is a very strong
argument in favor of both MSSM and Grand Unification. In theories with
large extra dimensions, this argument is apparently lost, as gravity
becomes strong at relatively low energies, so completely new physics
(strings) is to set in many orders of magnitude below $M_{GUT}$.

The situation is not so hopeless, however.
There have been discussed
at least two possibilities to obtain gauge coupling
unification in theories with large extra dimensions. One of them
exploits power-law running of couplings in higher-dimensional theories
 \cite{DDG}, another invokes massless fields propagating in two
large transverse dimensions 
\cite{Bachas:1998kr,Antoniadis:1999ax,Arkani-Hamed:1999yp} and leads
to
logarithmic unification.

The idea of power-law unification is as follows.
Suppose that
there
exists an energy scale $\mu_0$ at which extra dimensions open up for
gauge, Higgs and possibly quark and lepton fields of MSSM. In other
words, suppose that MSSM particles can leave our brane provided they
have energy exceeding $\mu_0$. Then below this scale, MSSM is
effectively four-dimensional, whereas above this scale it is
$D$-dimensional. In four-dimensional language, each MSSM particle has
its KK partners whose masses start at $\mu_0$. The scale fundamental
gravity scale $M$
should then be considered as an ultraviolet cut-off for the
$D$-dimensional MSSM. Since MSSM is non-renormalizable in more than
four dimensions, the low energy values of the gauge couplings depend
strongly (as a power law) on $M$ and $\mu_0$. In this sense
the gauge couplings exhibit power-law running at  scales
above $\mu_0$. The question is
whether there exist $M$ and $\mu_0$, both very roughly in the
TeV range, such that the gauge coulings unify at the cut-off scale,
\be
  \alpha_1^{-1} (M) =
 \alpha_2^{-1} (M) =
 \alpha_3^{-1} (M) \equiv \alpha_{GUT'}
\label{e3*}
\ee
and yet their low energy values are equal to the experimentally
measured ones,
\begin{eqnarray}
   \alpha_Y^{-1}(M_Z) &=& 98.3 \; , \;\;\; \mbox{i.e.}\; , \;\;
\alpha^{-1}_{1} (M_Z) = \frac{3}{5} \alpha_Y (M_Z) = 59.0
\nonumber \\
   \alpha_2^{-1}(M_Z) &=& 29.6
\nonumber \\
   \alpha_3^{-1}(M_Z) &=& 8.5
\end{eqnarray}
The relation (\ref{e3*}) would then substitute the conventional
unification of couplings and suggest that there is a Grand Unified
Theory above the fundamental scale
$M$. It is the {\it power-law}
 running of couplings that
makes this possibility not inconceivable.

Let us consider a theory in $d$ flat extra dimensions compactified on
a torus, as we did before in this Section. Consistent embedding of
MSSM in a higher-dimensional theory requires additional particles
above the scale $\mu_0$. A minimal extension \cite{DDG} is that the KK
tower has effective $N=2$ supersymmetry: from the four-dimensional
viewpoint,
there is $N=2$ vector
supermultiplet for each gauge group, an $N=2$ hypermultiplet
for the two Higgs fields and $\eta$ families of $N=2$ hypermultiplets
of quarks and leptons. With this matter content, the one-loop relation
between the low energy values of the gauge couplings and their values
at the cut-off is \cite{DDG}
\be
\alpha^{-1}_i (M) = \alpha^{-1}_i (M_Z) - 
\frac{b_i}{2\pi} \ln \frac{M}{M_Z} +
\frac{\tilde{b}_i}{2\pi} \ln \frac{M}{\mu_0}
-\frac{\tilde{b}_i \pi^{\frac{d}{2} -1}}{d^2 \Gamma (\frac{d}{2})}
\left[ \left(\frac{M}{\mu_0}\right)^d - 1 \right]
\label{e4*}
\ee
where $(b_1, b_2, b_3) = (33/5 ,1, -3)$ are the usual MSSM
one-loop $\beta$-function coefficients, and 
\[
  (\tilde{b}_1,  \tilde{b}_2, \tilde{b}_3) =
\left( \frac{3}{5} + 4\eta , -3 + 4\eta , -6 + 4 \eta \right)
\]
are ``$\beta$-function coefficients'' of higher-dimensional theory.
The power-law dependence of 
$[\alpha_i^{-1}(M_Z) - \alpha_i^{-1} (M)]$ on $M$ and
$\mu_0$, which is evident from eq.~(\ref{e4*}), is just the power-law
running of couplings; it reflects their dimensionality in $D=d+4$
dimensions.

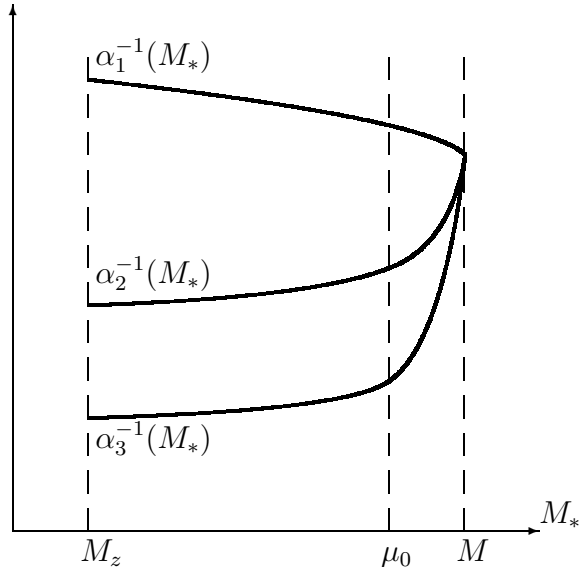
\begin{figure}[!hb]
\begin{center}
\unitlength1mm
\begin{picture}(70,70)(0,0)
\put(0,0){\vector(1,0){70}}
\put(0,0){\vector(0,1){70}}
\put(70,1){$M_{*}$}
\put(9,-4){$M_{z}$}
\multiput(10,0)(0,5){13}{\line(0,1){3}}
\put(49,-4){$\mu_{0}$}
\multiput(50,0)(0,5){13}{\line(0,1){3}}
\put(59,-4){$M$}
\multiput(60,0)(0,5){13}{\line(0,1){3}}
\linethickness{1pt}
\put(11,11){$\alpha_{3}^{-1}(M_{*})$}
\qbezier(10,15)(45,16)(50,20)\qbezier[500](50,20)(57,25)(60,50)
\put(11,33){$\alpha_{2}^{-1}(M_{*})$}
\qbezier(10,30)(40,31)(50,35)\qbezier[300](50,35)(58,38)(60,50)
\put(11,62){$\alpha_{1}^{-1}(M_{*})$}
\qbezier(10,60)(55,55)(60,50)
\end{picture}
\caption{\label{6}Schematic plot of the unification of
gauge couplings in the
presence of extra dimensions. For fixed $\alpha_i (M_z)$, the gauge
couplings at the cut-off scale $M_{*}$ depend on this scale
(logarithmically at $M_{*} < \mu_0$ and as power law at
  $M_{*} > \mu_0$). Unification occurs at $M_{*} = M$, where $M$ is
interpreted as the fundamental scale.}
\end{center}
\end{figure}

Somewhat miraculously, the unification of couplings indeed takes place
\cite{DDG}, as shown schematically in Fig. 6.
This occurs irrespectively of the number of extra dimensions, $d$, and 
the number of bulk quark-lepton generations, $\eta$. Namely, for given
$M$ (and fixed $d$ and $\eta$), the parameter $\mu_0$ can be
chosen
in such a way that eq.~(\ref{e3*}) is satisfied. Like in the
four-dimensional MSSM, this property is non-trivial in
higher-dimensional theories: by choosing one parameter $\mu_0$ one
satisfies two equations simultaneously. 
The unification
occurs for $\mu_0$ just (within  about an order of magnitude) below
$M$ --- the couplings run very fast above $\mu_0$. The gauge
coupling at the unification point is safely smaller than unity,
\[
  \alpha_i (M) \equiv \alpha_{GUT'} \ll 1
\]
It has been argued in Ref.~\cite{DDG} that this picture is unchanged
when higher loops are taken into account. Thus, contrary to the naive
expectation, the property of coupling constant unification
may be inherent in theories with large extra dimensions, although
instead of logarithmic unification, one may have faster, power-law
unification in these theories.

It is worth noting that in theories with large extra dimensions,
unification of gauge couplings does not, generally
speaking, require supersymmetry. In a number of non-supersymmetric
higher-dimensional extensions
of the Standard Model, gauge couplings exhibit power-law unification
as well \cite{DDG}.  
Hence, both motivations for supersymmetry --- stabilization
of the electroweak scale and gauge coupling unification --- are not
as strong in theories with large extra dimensions as they are
in
four-dimensional theories.


The  power-law unification has its problems, however. A technical
problem is that this picture is generically unstable against the
effects of possible non-renormalizable operators added at the cut-off
scale. Aesthetically, it is not appealing that the ``old'' unification
is given up: the logarithmic unification inherent in MSSM appears a
pure accident.

The logarithmic ``running'' of gauge couplings in theories with large
extra dimensions may occur in the following, rather unexpected way
\cite{Bachas:1998kr,Antoniadis:1999ax,Arkani-Hamed:1999yp}. Suppose,
in the spirit of string theory,  that the gauge couplings are in fact
massless scalar fields $\alpha_i^{-1}(z)$ which propagate  in
two large extra dimensions\footnote{The actual number of large extra
dimensions need not be equal to two; it is important only that the scalar
fields are effectively two-dimensional \cite{Arkani-Hamed:1999yp}.}.
Suppose also that  besides our brane, there exists another brane which
serves as a source for these fields, and at
which the couplings are unified,
\[
  \alpha_i^{-1} (z_{source}) = \alpha_0^{-1} \; , \;\;\; i=1,2,3
\]
 Naturally, the distance between our
and source branes is of the order of the size of extra dimensions $R$.
Since the two-dimensional propagator is a logarithm, the values of
$\alpha_i^{-1} (z)$ at the position of our brane are\footnote{Here,
the energy scale for $\alpha^{-1}_i (z_{our})$ and $\alpha^{-1}_0$
is the fundamental scale $M$.}
\be
  \alpha^{-1}_i (z_{our}) = \alpha^{-1}_0 - 
\frac{c_i}{2\pi} \ln~(|z_{our} - z_{source}| M)
\label{MR2*}
\ee
with some constants $c_i$,
where the fundamental scale $M$ appears in the logarithm on
dimensional grounds. For $|z_{our} - z_{source}| \sim R \sim
M_{pl}/M^2$
(as  given by eq.~(\ref{26*}) with $d=2$) the logarithmic term here is
\be
- \frac{c_i}{2\pi} \ln \left(\frac{M_{Pl}}{M}\right)
\label{MR2**}
\ee
which is almost the same logarithm as $ \ln
\left(\frac{M_{GUT}}{M_{EW}}\right)$
that occurs in conventional Grand Unification. 

Unlike conventional renormalization group evolution,  ``running''
described by 
eq.~(\ref{MR2*}) is an infrared effect. Still, it works in right
direction: the values of gauge couplings in our world differ from the
``fundamental'' value $\alpha_0^{-1}$ by  large logarithms. The most
miraculous thing is that in certain stringy brane constructions,
constants in eq.~(\ref{MR2*}) are precisely the $\beta$-function
coefficients of the theory localizad on our brane! The discussion of
this property and its relation to string dualities goes far beyond
this mini-review; the reader may consult
Ref.~\cite{Arkani-Hamed:1999yp}
and references therein. We stress only that this property opens up a
way towards gauge coupling {\it unification}, not merely logarithmic
``running'' of couplings. Although a phenomenologically acceptable
model with MSSM on our brane has not  been constructed along these
lines yet, this mechanism is certainly promising, as it suggests that
the logarithmic unification existing in MSSM may have its close
counetrpart in theories with large extra dimensions.


\subsection{Problem with proton stability}

The above discussion brings us to one more problem with large (and
also infinite) extra dimensions --- potentially too fast proton
decay. This problem is apparent if one takes the attitude of the
previous subsection and considers the unification of gauge couplings
as a signal of Grand Unification of strong and electroweak
interactions at the scale $M$. In conventional Grand Unified
Theories, dimension six operators lead to proton decay with life-time
of order\footnote{In supersymmetric GUTs, proton decay occurs also due
to dimension five operators. We consider the contribution (\ref{e6*})
merely as an illustration.}
\be
     \tau_p \sim \frac{1}{\alpha_{GUT}^2 M_p} 
\left( \frac{M_{GUT}}{M_p} \right)^4
\label{e6*}
\ee
With $M_{GUT} \sim 10^{16}$~GeV, this is consistent with experimental
limits on the proton instability, but if $M_{GUT}$ is shifted to
$M$ (which is roughly in the TeV range in ADD scenario discussed
in this Section), this estimate gives way too small proton life-time. 
Grand Unification at the scale $M$ should somehow occur in such
a way that the proton width is suppressed by many orders of magnitude,
as compared to the dimensional estimate.


More generally,  global
quantum numbers, such as baryon and lepton number, may not be
conserved
when quantum gravity effects are included. With fundamental gravity
scale $M_{Pl} \sim 10^{19}$~GeV, dimensional estimate similar to
eq.~(\ref{e6*}), i.e.,
\[
      \tau_p \sim \frac{1}{M_p}\left( \frac{M_{Pl}}{M_p} \right)^4
\]
shows that proton instability due to gravity effects is not
phenomenologically dangerous. However, if the fundamental gravity
scale is in the TeV range, the same estimate gives unacceptably small
proton life-time. 
Hence, one has to invoke special mechanisms (such as discrete gauge
symmetries
\cite{Krauss:1989zc,Alford:1990ch,Preskill:1990bm,Alford:1991pt,Kamionkowski:1992mf})
to forbid proton decay in theories with large extra dimensions.

\section{Non-factorizable geometry}

\subsection{Warped extra dimension}

Until now we have ignored the energy density of the brane itself,
i.e., the gravitational field that the brane produces. Here we shall
see that a gravitating brane induces an interesting geometry in 
multi-dimensional space, and that a number of novel properties
emerge.

When considering distance scales much larger than the brane thickness,
one may view the brane as a delta-function source of the gravitational
field. In the simplest case, the gravitating brane is characterized by
just one parameter, the energy density per unit three-volume 
$\sigma$. This quantity is also called brane tension.
We shall mostly discuss the case of one extra dimension, so
the five-dimensional gravitational action in the presence of the
brane is
\be
S_g = 
     -\frac{1}{16\pi G_{(5)}} \int~d^4x~dz~\sqrt{g^{(5)}} R^{(5)}
   - \Lambda  \int~d^4x~dz~\sqrt{g^{(5)}}
    -\sigma  \int~d^4 x~ \sqrt{g^{(4)}}
\label{51*}
\ee
where $\Lambda$ is the five-dimensional cosmological constant,
and the integral in the last term is evaluated along the world surface
of the 3-brane with $g^{(4)}_{\mu \nu}$ being the induced metric.

The resulting field equations are straightforward to obtain. In the
bulk, these are the standard five-dimensional Einstein equations with
the cosmological constant $\Lambda$, while the last term in
eq.~(\ref{51*}) gives rise to the Israel junction conditions 
\cite{Israel:1966rt} on the brane surface (for pedagogical
presentation of the Israel conditions see, e.g.,
Ref.~\cite{Berezin:1987bc}). Notably, this set of equations allows for
a solution preserving four-dimensional Poincar\'e invariance. This fact
was extensively discussed in the $D$-brane context (see, e.g,
Ref.~\cite{Freedman:2000gk} and references therein)
and its relevance for phenomenological models has been stressed in 
Refs.~\cite{Gogberashvili:2000iu,Gogberashvili:1999tb,RS1,RS2}. The
existence of four-dimensionally flat solution requires fine-tuning
between $\Lambda$ and $\sigma$: the five-dimensional cosmological
constant must be negative and equal to \cite{RS1}
\be
 \Lambda = - \frac{4\pi}{3} G_{(5)} \sigma^2
\label{53*}
\ee
(note that the parameters here have dimensions $[\Lambda] = M^5$,
$[\sigma] = M^4$, $[G_{(5)}] = M^{-3}$). This fine-tuning is very
similar to fine-tuning of the cosmological constant to zero
in conventional
four-dimensional gravity; indeed, if eq.~(\ref{53*}) does not hold,
the intrinsic geometry on the brane
is (anti-)de Sitter rather than flat.

With the relation (\ref{53*}) satisfied, the four-dimensionally flat
solution has the form \cite{RS1}
\be
   ds^2 = a^2(z) \eta_{\mu \nu} dx^{\mu} dx^{\nu} - dz^2
\label{RS}
\ee
where  $\eta_{\mu \nu}$ is the four-dimensional Minkowski metric and
the ``warp factor'' has the form
\[
   a(z) = \mbox{e}^{-k |z|}
\]
where
\be
   k = \frac{4\pi}{3} G_{(5)} \sigma
\label{p17*}
\ee
The brane is located at $z=0$.

To see that this is indeed a solution to the complete system of the
Einstein equations, we write for the metric (\ref{RS})
\begin{eqnarray}
  G_{\mu \nu} \equiv R_{\mu \nu}^{(5)} 
- \frac{1}{2} g_{\mu \nu}^{(5)} R^{(5)} &=&
g^{(5)}_{\mu \nu} 
\left[ -3 \frac{a''}{a} -3 \left(\frac{a'}{a} \right)^2 \right]
\nonumber \\
 G_{z \mu} &=& 0
\nonumber \\
 G_{zz} &=&  g^{(5)}_{zz} \left[-6 \left(\frac{a'}{a}\right)^2\right]
\end{eqnarray}
where prime denotes the derivative with respect to $z$.
In is then straightforward to see that metric (\ref{RS}) is a solution
to the Einstein equations
\begin{eqnarray}
  G_{\mu \nu} &=& 8 \pi G_{(5)} \Lambda g_{\mu \nu}^{(5)}
+  8 \pi G_{(5)} \sigma  g_{\mu \nu}^{(5)} \delta (z)
\nonumber \\
 G_{z \mu} &=& 0
\nonumber \\
 G_{zz} &=& 8 \pi G_{(5)} \Lambda g_{zz}^{(5)}
\end{eqnarray}
provided eq.~(\ref{p17*}) is satisfied, and
\[
  k^2 = - \frac{4\pi}{3}  G_{(5)} \Lambda
\]
which is equivalent to eq.~(\ref{53*}). Note that eq.~(\ref{p17*})
comes from the requirement
\[
   - 3 \frac{[a']}{a}  g_{\mu \nu}^{(5)}
=  8 \pi G_{(5)} \sigma  g_{\mu \nu}^{(5)} \; , \;\;\; \; z=0
\]
where $[a']$ denotes the jump of $a'$ at $z=0$. The latter requirement
is essentially the Israel condition in the case considered.

The metric (\ref{RS}) is non-factorizable: unlike the metrics
appearing in the usual Kaluza--Klein scenarios, it does not correspond
to a
product of the four-dimensional Minkowski space and a (compact)
manifold of extra dimensions. This metric rather corresponds to two
patches of anti-de Sitter space of radius $1/k$ glued together along
$z=0$, i.e., along the brane. The four-dimensional hypersurfaces
$z=\mbox{const}$ are flat; in particular, the metric induced on the
brane is the Minkowski metric $\eta_{\mu \nu}$. 

At this point it is worth mentioning one
property of the metric (\ref{RS}). Due to four-dimensional Poincar\'e
invariance, every field in this background can be decomposed into
four-dimensional plane waves,
\[
   \phi \propto \mbox{e}^{i p_{\mu} x^{\mu}} \phi_p (z)
\]
The coordinate four-momentum $p_{\mu}$ coincides with the physical
momentum
{\it on the brane}, but from the point of view of an observer residing
at $z \neq 0$, the physical four-momentum is larger,
\be
  p_{\mu}^{phys} (z)
= \frac{1}{a(z)} p_{\mu} = \mbox{e}^{k|z|} p_{\mu}
\label{55*}
\ee
The modes which are soft on the brane become harder away from the
brane. This scaling property is behind many peculiarities of physics
in the background (\ref{RS}).

\subsection{Two-brane set up}

There are several approaches which make use of the solution
(\ref{RS}). One of them \cite{RS1} is to make extra dimension compact by
introducing two branes: one with positive tension $\sigma$ at $z=0$,
and the other with negative tension $(-\sigma)$ located at distance
$z_c$, see Fig. 7.
Allowing the negative tension  brane  to vibrate freely is
dangerous, as this will give rise to physical excitations of
arbitrarily large negative energy (see Ref.~\cite{Pilo} for detailed
discussion). To circumvent this problem, the branes are placed at
fixed points of an orbifold; in our case, this means that all bosonic
fields, including gravity, are required to be symmetric under
reflections with respect to both $z_c$, the position of the negative
tension brane, and $z=0$, the position of positive tension one
(fermion fields may have more complicated symmetry properties).
The metric (\ref{RS}) is still a solution of the complete set of
Einstein equations in the presence of the two branes; extra dimension
is compact, as the coordinate
$z$ runs now from $z=0$ to $z=z_c$,
\[
   z \in [0, z_c]
\]
The orbifold boundary conditions (reflection symmetry) project the
undesirable negative energy modes out, and there remain positive
energy excitations only.

\begin{figure}[!hb]
\begin{center}
\unitlength1mm
\begin{picture}(50,65)(0,-15)
\put(10,26){$a(z)$}
\put(1,-4){$z=0$}
\put(38,-4){$z=z_{c}$}
\put(-2,-14){$+\sigma$}
\put(48,-14){$-\sigma$}
\qbezier(0,45)(10,10)(50,3)
\put(0,0){\vector(1,0){50}}
\linethickness{3pt}
\put(0,-10){\line(0,1){60}}
\put(50,-10){\line(0,1){60}}
\end{picture}
\caption{\label{7}Two-brane set up: 
branes of positive and negative tensions and
warp factor between them.}
\end{center}
\end{figure}
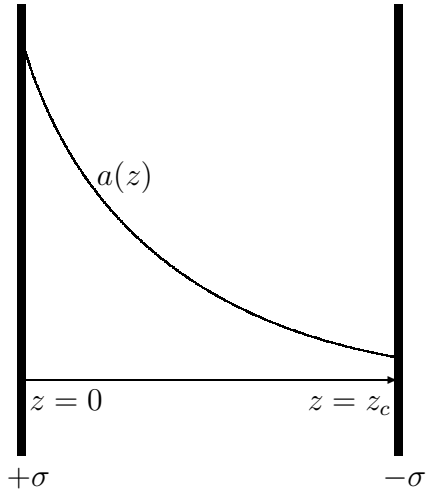

Let us consider small perturbations about the metric (\ref{RS}). 
To make the long story short, one can use the gauge
\[
   g_{55} = -1\; , \;\;\;\; g_{5\mu} = 0
\]
i.e., consider perturbed metric of the form
\[
 ds^2 = [a^2(z) \eta_{\mu \nu} + h_{\mu \nu}(x,z)] dx^{\mu} dx^{\nu} - dz^2
\]
If there are no sources of the gravitational field except for the bulk
cosmological constant and the two branes, one can further specify the
coordinate frame (i.e., fix the gauge), so that $h_{\mu\nu}$ are
transverse and trace-free in the bulk,
\[
   \partial_{\mu} h^{\mu}_{\nu} = 0 \; , \;\;\;\;
   h_{\mu}^{\mu} = 0
\]
For all types of perturbations but one, this frame is at the same time
Gaussian normal with respect to both branes. This means that the
positions of the branes are still $z=0$ and $z=z_c$. Then all
components of $h_{\mu\nu}$ obey the same equation (we omit the
subscripts) 
\be
    h^{\prime \prime} - 4 k^2 h - \frac{m^2}{a^2(z)} h = 0
\label{59*}
\ee
where
\[
   m^2 = \eta^{\mu \nu} p_{\mu} p_{\nu} 
\]
is the four-dimensional mass of the perturbation. The junction
conditions on the branes are (assuming the orbifold symmetry)
\be
   h' + 2kh = 0 \; \;\;\;\; \mbox{at} \; z=0 \;\; \mbox{and} \; z=z_c
\label{59**}
\ee
Equations (\ref{59*}) and (\ref{59**}) determine the mass spectrum of
the KK gravitons, where the mass is defined with respect to the
positive tension brane, cf. eq.~(\ref{55*}).

Before considering this spectrum, let us point out that there exists
one scalar mode which cannot be treated in the above way. This mode
--- the radion --- is massless and corresponds to oscillations of the
relative distance between the two branes. Its properties are
discussed, e.g., in Refs.~\cite{Lisa,CGR,Pilo}. In many
phenomenological
models
based on this set up, massless radion is unacceptable (we shall
briefly discuss this point below). Giving the radion a mass
corresponds to stabilizing the distance between the branes;  field
theory mechanisms for this stabilization is suggested, e.g., in 
Ref.~\cite{Goldberger:1999uk,Luty:2000cz}. We shall not consider the radion mode
in what follows, assuming that the distance between the branes is
stabilized in one or another way.

Let us now turn to the graviton spectrum, i.e., solutions 
to eqs~(\ref{59*}) and (\ref{59**}).
There exists a zero mode, $m^2=0$, whose wave function, up to
normalization, is
\be
  h_0 (z) = \mbox{e}^{-kz}
\label{61*}
\ee
This mode describes the usual four-dimensional gravity. Unlike in the
Kaluza--Klein theories with factorizable geometry, the zero-mode wave
function depends on $z$ non-trivially, and decreases towards
$z=z_c$. This suggests that the gravitational coupling between
particles residing on the negative tension brane is weak as compared
to the positive tension brane. We shall discuss this feature in more
detail later on. 

Solutions to eq.~(\ref{59*}) obeying the boundary condition
(\ref{59**}) at $z=0$ (not yet at $z=z_c$) are, again up to
normalization, 
\be
   h_m (z) = N_1\left(\frac{m}{k}\right) 
J_2\left(\frac{m}{k} \mbox{e}^{kz}\right) -
J_1\left(\frac{m}{k}\right) 
N_2\left(\frac{m}{k} \mbox{e}^{kz}\right)  
\label{62**}
\ee
where $N$ and $J$ are the Bessel functions. The mass spectrum is
determined by the boundary condition (\ref{59**}) at $z=z_c$. One
obtains that the mass splitting between the KK modes is of order
\be
  \Delta m \sim k\mbox{e}^{-kz_c}
\label{62*}
\ee
The phenomenological interpretation of these results depends on
whether the Standard Model particles are bound to the brane of
positive or negative tension.

\subsection{Matter on negative tension brane and hierarchy}

Let us first consider the possibility that the ordinary matter resides
on the negative tension brane (RS1 scenario \cite{RS1}). We are interested in
gravitational interactions of this matter at large distances; then the
dominant contribution to the gravitational attraction is due to the
zero graviton mode. It is
convenient to rescale the four-dimensional coordinates in such a way
that the warp factor is equal to 1 at the negative tension brane
(i.e., at $z=z_c$),
\[
   a(z) = \mbox{e}^{k(z_c - z)}
\]
Similarly, it is convenient to normalize the zero mode so that it is
equal to 1 at the negative tension brane. Then
the massless gravitational perturbations are described by the field
\be
  h_{\mu \nu} (x,z) = \mbox{e}^{2k(z_c - z)}  h_{\mu \nu}^{(4)} (x)
\label{64*}
\ee
The coordinates $x^{\mu}$ are now the physical coordinates on the
negative tension brane, and the four-dimensional graviton 
field $ h_{\mu \nu}^{(4)} (x)$ couples
to energy-momentum of the ordinary matter in the usual way,
\be
  S_{int} = \int~d^4x~  h_{\mu \nu}^{(4)} T^{\mu \nu}
\label{64+}
\ee
The strength of gravitational interactions is read off from the
quadratic part of the action for $h_{\mu \nu}^{(4)}$. This is obtained
by plugging the expression (\ref{64*}) in the five-dimensional
gravitational action (\ref{51*}). Schematically, one has
\begin{eqnarray}
S_g &=& 
\frac{1}{16\pi G_{(5)}} 
\int_{-z_c}^{z_c}~\frac{dz}{a^2(z)}~d^4x~(\partial_{\mu} h)^2
 \nonumber \\
&=& 
\frac{1}{8\pi G_{(5)}} 
\int_{0}^{z_c}~dz~\mbox{e}^{2k(z_c -z)}~\int~d^4x~(\partial_{\mu} h^{(4)})^2
\nonumber \\
&=&
\frac{(\mbox{e}^{2kz_c} -1)}{16\pi G_{(5)} k}~
\int~d^4x~(\partial_{\mu} h^{(4)})^2
\label{65*}
\end{eqnarray}
(the factor $a^{-2}$ in the first integral can be either obtained
directly, or inferred from the structure of eq.~(\ref{59*})).
The integral in the last expression is the quadratic action
for four-dimensional gravitons. Hence, the four-dimensional Newton's
constant is
\be
G_{(4)} = G_{(5)}k \frac{1}{\mbox{e}^{2kz_c} -1}
\label{65**}
\ee
which means that at relatively large $z_c$, the gravitational
interactions of matter residing on the negative tension brane are
weak.

This observation opens up a novel possibility to address the hierarchy
problem. Indeed, one can take the fundamental five-dimensional gravity
scale, as well as the inverse anti-de Sitter radius $k$ to be of the
order of the weak scale, $M_{EW} \sim 1$~TeV. As is clear from
eq.~(\ref{65**}), the effective four-dimensional Planck mass is then of
order
\be
   M_{Pl} \sim \mbox{e}^{kz} M_{EW}
\label{66*}
\ee
which means the exponential hierarchy between the Planck mass and the
weak scale: for $z_c$ only about 37 times larger than the anti-de
Sitter radius $k^{-1}$, the value of $M_{Pl}/M_{EW}$ is of the right
order of magnitude. 

One may wonder whether the low energy theory of gravity obtained in
this set up is indeed the conventional four-dimensional
General Relativity. It has been shown explicitly in
Refs.~\cite{Garriga:2000yh,Giddings:2000mu} 
that this is indeed the case,
{\it provided} the radion is given a mass (see also
\cite{Lisa}). Otherwise the radion would act as a Brans--Dicke field
with unacceptably strong (TeV scale) coupling to matter.

Let us now turn to KK gravitons. The mass splitting (\ref{62*}) refers
to the masses measured by an observer on the positive tension
brane. According to eq.~(\ref{55*}), the physical masses measured by
an observer on the negative tension brane are of order
\[
m_{grav} \sim k
\]
Hence, KK gravitons have masses in TeV range, in clear distinction to
ADD scenario. In particular, the cosmological difficulties inherent in
ADD picture do not appear here: the maximum temperature of the
Universe is allowed to be just below the TeV range.

Unlike the zero mode, the coupling of KK gravitons to matter residing
on the negative tension brane is characterized by the fundamental mass
scale (of order $M_{EW}$). To see this, we write the massive modes as
follows (for $m$ somewhat larger than $k\mbox{e}^{-kz_c}$)
\be 
h_{m}(x,z) =
\mbox{e}^{k(z_c - z)/2} \sin \left(\frac{m}{k} \mbox{e}^{kz} 
- \varphi_m \right) h_m^{(4)} (x)
\label{68*}
\ee
This expression is valid at $(m/k) \mbox{e}^{kz} \gg 1$, while the KK
wave functions decrease towards $z =0 $. The pre-factor in
eq.~(\ref{68*}) has been chosen in such a way that the
four-dimensional fields $ h_m^{(4)} (x)$ couple to matter at $z=z_c$
with unit strength, cf. eq.~(\ref{64+}). In the same way as
eq.~(\ref{65*}) one obtains the quadratic action
\begin{eqnarray}
S_{g,m} &=& 
\frac{1}{16\pi G_{(5)}} 
\int~\frac{dz}{a^2(z)}~d^4x~(\partial_{\mu} h_m)^2
 \nonumber \\
&=& 
\frac{1}{8\pi G_{(5)}} 
\int_{0}^{z_c}~dz~\mbox{e}^{-k(z_c -z)}~\int~d^4x~(\partial_{\mu} h^{(4)}_m)^2
\nonumber \\
&=&
\frac{(1- \mbox{e}^{-kz_c} )}{8\pi G_{(5)} k}~
\int~d^4x~(\partial_{\mu} h^{(4)}_m)^2
\label{68**}
\end{eqnarray}
Hence, the mass scale determining the interactions of KK gravitons
with matter is of order
\[
  M_m \sim \frac{1}{\sqrt{G_{(5)}k}}
\]
which is of order $M_{EW}$. The interaction of matter and KK gravitons
becomes strong in the TeV energy range.

Thus, RS1 scenario leads to exponential hierarchy between the weak and
Planck scales. Similarly to ADD, gravity becomes strong at TeV
energies; manifestations of this phenomenon in collider experiments
will be quite different in RS1 model as compared to ADD 
\cite{Davoudiasl:2000jd}. The reason is of course that the graviton
spectra are entirely different; a distinctive feature of RS1 collider
phenomenology is TeV scale graviton resonances quite strongly coupled to
ordinary particles. 
For further discussion of RS1 phenomenology see
Refs.~\cite{Arkani-Hamed:2000ds} and references therein.

\subsection{Matter on positive tension brane}

Another option is that the conventional matter resides on the positive
tension brane. The analysis similar to that leading to eq.~(\ref{65**})
shows that in this case the effective four-dimensional Newton's
constant is 
\be
G_{(4)} = G_{(5)}k \frac{1}{1 - \mbox{e}^{-2kz_c} }
\label{70*}
\ee
If one does not introduce huge hierarchy between the fundamental
five-dimensional gravity scale and the inverse anti-de Sitter radius
$k$, the fundamental scale must be of order of $M_{Pl}$. This does not
mean, however, that the exponential hierarchy between the scales
cannot be generated. There is a possibility that the electroweak
symmetry breaking and/or supersymmetry breaking occur due to physics
on the negative tension brane, and are transfered to ``our'' brane by
one or another mechanism \cite{Gherghetta:2000kr}. In that case the
electroweak scale and/or supersymmetry breaking scale in our world are
naturally exponentially smaller than the Planck scale, essentially
because of the scaling relation (\ref{55*}), and the exponential
hierarchy (\ref{66*}) is again generated. Concerete models of this
sort have been discussed in Ref.~\cite{Gherghetta:2000kr}, and their
phenomenology turned out to be quite interesting. Note that the masses
of KK gravitons are again in the TeV range: these are given by
eq.~(\ref{62*}) with $k \sim M_{Pl}$.

\section{Infinite extra dimension}

\subsection{Localized graviton}

The graviton zero mode (\ref{61*}) that appeared in the set up
of Section 5, is normalizable for $z_c \to \infty$, i.e., for negative
tension brane moved away. This means that gravity is still
localized if
there exists a single positive tension brane only, and extra dimension
is infinite \cite{RS2}. Hence, one is lead to consider a set up,
called RS2 \cite{RS2}, with matter residing on the positive tension
brane and experiencing four-dimensional gravity law at large distances
due to the exchange of the graviton zero mode. The fact that gravity
is four-dimensional at large distances is clear also from
eq.~(\ref{70*}): in the limit $z_c \to \infty$, the four-dimensional
Newton's constant tends to a finite value
\[
   G_{(4)} = G_{(5)} k
\]
Obviously, in this simplest set up, unlike in constructions of
Sections 4 and 5, the hierarchy between the Planck and weak scales is
not explained by physics of extra dimension, and one has to rely upon
other, more conventional mechanisms (we shall mention another option
later on). The interest in this simplest set up with infinite extra
dimension is, on the one hand, due to potentially interesting physics
at {\it low} energies, and, on the other hand, due to its connection
to adS/CFT correspondence (for a brief review of adS/CFT see, 
e.g., Ref.~\cite{Akhmedov:1999rc}). 

One property of the anti-de Sitter geometry in the bulk is worth
mentioning. Although the distance from the brane to $z=\infty$,
measured along the $z$-axis is infinite, it is straightforward to see
that $z=\infty$ is in fact a particle horizon. Indeed, let us consider, as an
example,  a massive particle that starts 
from the brane at $t=0$ and ${\bf x} =0$ with zero velocity, and then
freely moves along the $z$-axis. The corresponding solution to the
geodesic equation is \cite{Muck:2000bb,Gregory:2000rh}
\[
  z_c (t) = \frac{1}{2k} \ln (1 + k^2 t^2)
\]
The particle accelerates towards $z \to \infty$, its velocity tends to
the speed of light. According to eq.~(\ref{RS}), the proper-time
interval is determined by
\[
  d\tau^2 = a^2(z_c(t)) dt^2 - \left( \frac{dz_c}{dt} \right)^2 dt^2
\]
The particle reaches $z=\infty$ at infinite time $t$, but finite
proper time
\[
  \tau = \int_0^{\infty} \frac{dt}{1+ k^2 t^2} = \frac{\pi}{2k}
\]
Hence, $z=\infty$ is indeed the particle horizon. When considering
physics in the background (\ref{RS}), one has to impose certain
boundary conditions at the horizon $z=\infty$ (which in principle may
affect physics on the brane); it is usually assumed that nothing comes
in from ``behind the horizon''.

To substantiate the claim that gravity experienced by matter residing
on the (positive tension) brane is effectively four-dimensional at
large distances, let us consider the Kaluza--Klein
gravitons. According to eq.~(\ref{62**}), the spectrum of KK gravitons
 {\it is continuous} and {\it starts from zero} $m^2$. In this
situation, the wave functions of KK gravitons are to be normalized to
delta-function (again with the measure $a^{-2}~dz$, see
eq.~(\ref{65*})),
\[
 \int~\frac{dz}{a^2(z)}~ h_m(z) h_{m'} (z) = \delta (m - m')
\]
Making use of the asymptotics of the Bessel functions, one obtains
that the properly normalized KK graviton wave functions are
\be
   h_m (z) = \sqrt{\frac{m}{k}} 
\frac{J_1\left(\frac{m}{k}\right) N_2\left( \frac{m}{k} \mbox{e}^{kz}\right)
- N_1\left(\frac{m}{k}\right) 
J_2\left( \frac{m}{k}
\mbox{e}^{kz}\right)}{\sqrt{\left[J_1\left(\frac{m}{k}\right)\right]^2 +
\left[N_1\left(\frac{m}{k}\right)\right]^2}}
\label{74*}
\ee
At large $z$, these wave functions oscillate,
\[
   h_m (z) = \mbox{const} \cdot \sin \left(\frac{m}{k} \mbox{e}^{kz}
+ \varphi_m \right) 
\]
whereas they decrease towards small $z$ and are suppressed at $z=0$,
\[
  h_m (0) = \mbox{const} \cdot \sqrt{\frac{m}{k}}
\]
The wave functions (\ref{74*}) correspond to gravitons escaping into
extra dimension, i.e., towards $z \to \infty$ (or coming towards the
brane from $z = \infty$). The coupling of these KK gravitons to
matter, residing on the brane, is fairly weak at small $m$, so their
production at relatively low energies (and/or temperatures) is
unimportant (for details, see Ref.~\cite{Hebecker:2001nv} and
references therein). Likewise, the contribution of virtual
KK  gravitons into low energy processes is small.

As an example, let us consider the contribution of KK graviton
exchange into gravitational potential between two unit point masses
placed on the brane. Each KK graviton produces the potential of Yukawa
type, so the total contribution is
\begin{eqnarray}
  \Delta V_{KK} (r) &=& - G_{(5)} \int_0^{\infty}~dm~
|h_{m}(0)|^2 \frac{\mbox{e}^{-mr}}{r} 
\nonumber \\
&=& - \frac{G_{(5)}k}{r} \cdot \mbox{const} \cdot
 \int_0^{\infty}~\frac{m dm}{k^2} \mbox{e}^{-mr}
\nonumber \\
&=&
- \frac{G_{(4)}}{r} \cdot \frac{\mbox{const}}{k^2 r^2}
\label{75*}
\end{eqnarray}
Hence, the gravitational potential, including the contribution of the
graviton zero mode, is \cite{RS2}
\[
   V(r) = - \frac{G_{(4)}}{r} \left( 1 + \frac{\mbox{const}}{k^2 r^2}
   \right) 
\]
The correction to Newton's law has power law behaviour at large $r$,
in contrast to theories with compact extra dimensions where the 
corrections are suppressed exponentially at large distances. However,
this correction is negligible at distances exceeding the anti-de
Sitter radius $k^{-1}$. It has been explicitly
shown in Refs.~\cite{Garriga:2000yh,Giddings:2000mu}
that the tensor structure of the gravitational interactions at large
distances indeed corresponds to (the weak field limit of) the
four-dimensional General Relativity. Note that the radion excitation
is absent in RS2 set up.

We have already mentioned that in RS2 set up with one brane, extra
dimension does not help to solve the hierarchy problem. It was pointed
out, however, that modest extension of this set up leads to
exponential hierarchy even if extra dimension is infinite
\cite{Lykken:2000nb}. Instead of assuming that our matter is bound to
the ``central'' brane, one may introduce one more, ``probe'' brane
which is placed at a position $z_c$ in extra dimension and, for
simplicity, has zero tension. Metric (\ref{RS}) on this probe brane
still has four-dimensional Poincar\'e invariance. If our matter is put
on the probe brane, the exponential hierarchy (\ref{66*}) is generated
in much the same way as in RS1 set up discussed in Section 5.

\subsection{Escape into extra dimension}

If one or more extra dimensions are infinite, one naturally expects
that particles may eventually leave our brane and escape into extra
dimension. In RS2 set up, this process is certainly possible for
gravitons, as the excitation of a KK mode is interpreted precisely as
escape of a graviton towards $z \to \infty$. If other fields have bulk
modes, the corresponding particles may also leave our brane. As an
example, even in the absence of gravity, fermions bound to the brane
by the mechanism presented in subsection 3.1, are capable of leaving
the brane provided they are given enough energy. As we discussed in
subsection 3.1, this would show up as a process like 
$e^+ e^- \to \mbox{nothing}$ which would be possible at {\it high}
energies. 

A novelty of the bulk with anti-de Sitter metric is that such
processes become possible also at {\it low} energies \cite{DRT1}.
The ultimate reason is again the scaling property (\ref{55*}):
energies which are low if measured at the brane position, become high
if measured at large $z$. Low energy physics on the brane is high
energy physics away from the brane. 

Quantitatively, this feature is
manifest in a peculiar property of KK continuum for fields having bulk
modes: the continuum starts from {\it zero} $m^2$ irrespectively of
the dynamics near the brane. Suppose now, that in the absence of
gravity, a field has a bound state of a non-zero mass, whose wave
function concentrates near the brane and hence corresponds to a
four-dimensional particle. When gravity is turned on, this would-be
bound state becomes embedded in a continuum of KK modes, which
describe particles capable of escaping to $z \to \infty$. Hence, this
would-be bound state becomes quasi-localized (there are no true bound
states embedded in continuum, unless the potential is very contrived):
its energy obtains an imaginary part which determines (finite)
probability of tunneling to large $z$.
Particle on the brane becomes metastable against escape 
into extra dimension. 

To illustrate this fairly general phenomenon, let us consider real
scalar field in the presence of the brane, with the action
\be
  S_{\phi} = \int~d^4 x~dz~\sqrt{g}~
\left[ \frac{1}{2} g^{AB} \partial_A \phi \partial_B \phi -
\frac{1}{2} V(z) \phi^2 \right]
\label{p28*}
\ee
where $x^A = (x^{\mu}, z)$ are coordinates in five-dimensional
space-time. Effects of the brane are encoded in the potential $V(z)$,
which is assumed  to tend to a (possibly, non-zero) non-negative
constant as $z \to \infty$ (we asuume the orbifold symmetry $z \to -z$
for simplicity). 
If gravity is switched off, the field $\phi$ obeys
the Klein--Gordon equation 
\[
   -\partial_{\mu}^2 \phi + \partial_z^2 \phi - V(z) \phi = 0
\]
The spectrum of four-dimensional masses is determined by the potential
$V(z)$,
\be
  p_{\mu}p^{\mu} \phi \equiv m^2 \phi =
[- \partial_z^2 + V(z)] \phi
\label{80+}
\ee
An interesting case is when the operator on the right hand side of 
this equation has discrete eigenmodes which correspond to
 particles trapped to the brane. 
This situation is shown in Fig. 8.
The continuum starts
at $m^2 = V(\infty)$.

\begin{figure}[!hb]
\begin{center}
\unitlength1mm
\begin{picture}(160,100)(-80,-40)
\put(-80,0){\vector(1,0){160}}
\put(0,-40){\vector(0,1){100}}
\put(40,37){$V(z)$}
\put(40,10){$\Phi_{m}(z)$}
\put(78,-4){$z$}
\qbezier(-20,20)(-35,6)(-70,1)
\qbezier(-20,20)(0,40)(20,20)
\qbezier(20,20)(35,6)(70,1)
\linethickness{2pt}
\qbezier(-30,35)(-40,35)(-60,35)
\qbezier(-15,10)(-17,35)(-30,35)
\qbezier(-15,10)(-12,-30)(0,-30)
\qbezier(15,10)(12,-30)(0,-30)
\qbezier(15,10)(17,35)(30,35)
\qbezier(30,35)(40,35)(60,35)
\end{picture}
\caption{\label{8}Binding potential $V(z)$ and bound state with $m^2 \neq 0$
in the absence of the warp factor.}
\end{center}
\end{figure}
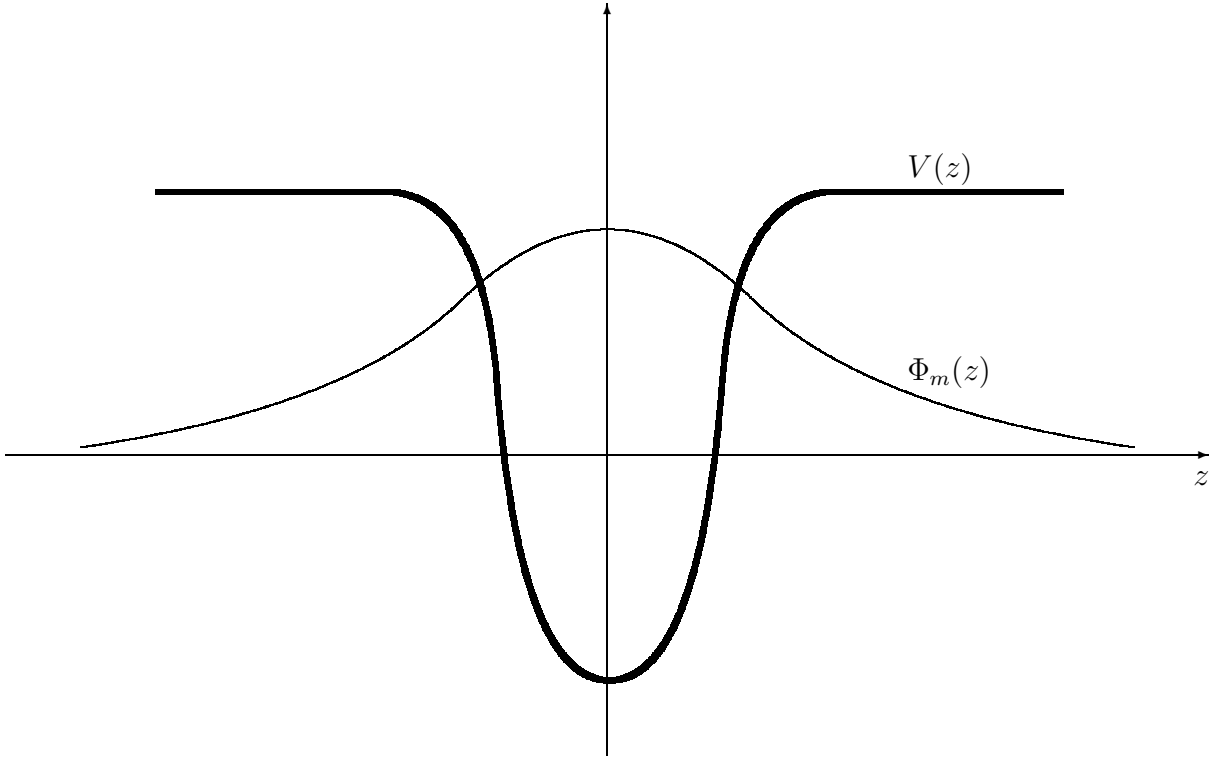

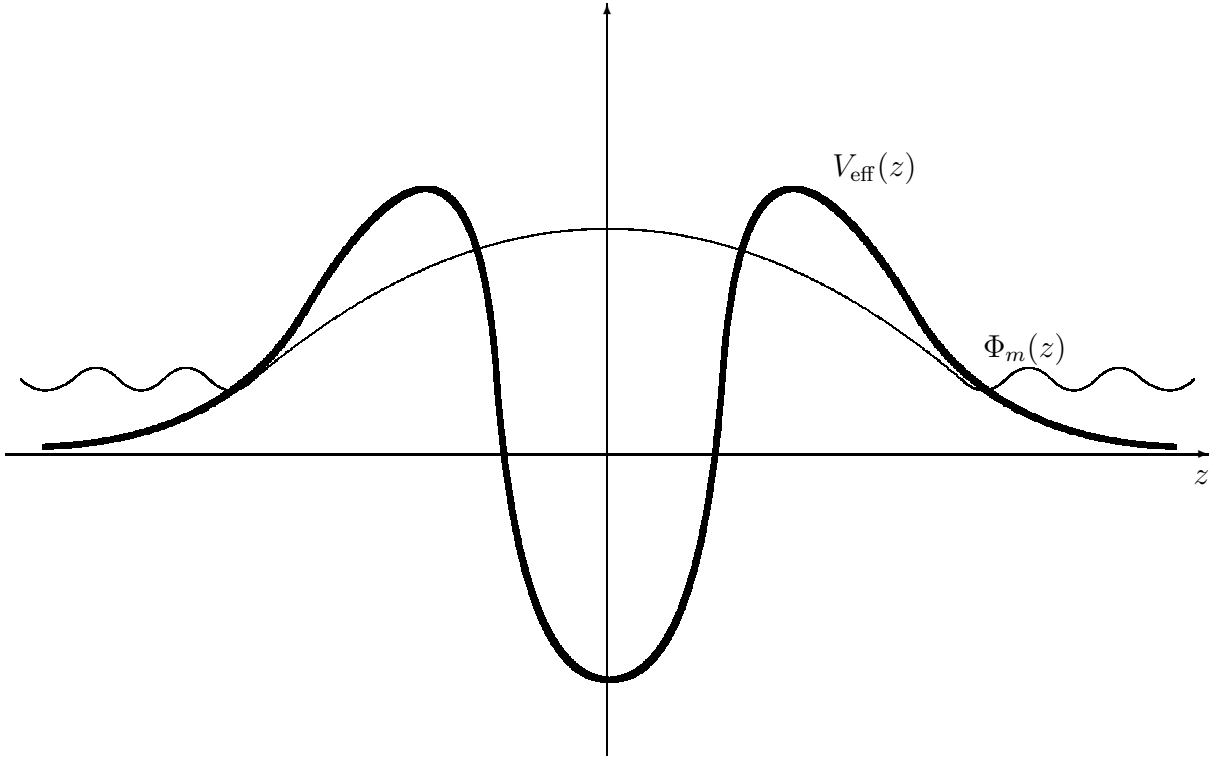
\begin{figure}[!hb]
\begin{center}
\unitlength1mm
\begin{picture}(160,100)(-80,-40)
\put(-80,0){\vector(1,0){160}}
\put(0,-40){\vector(0,1){100}}
\put(30,37){$V_{\mathrm{eff}}(z)$}
\put(50,13){$\Phi_{m}(z)$}
\put(78,-4){$z$}
\qbezier(-47,10)(0,50)(47,10)
\qbezier(47,10)(50,7)(53,10)\qbezier(53,10)(56,13)(59,10)
\qbezier(59,10)(62,7)(65,10)\qbezier(65,10)(68,13)(71,10)
\qbezier(71,10)(75,7)(78,10)
\qbezier(-47,10)(-50,7)(-53,10)\qbezier(-53,10)(-56,13)(-59,10)
\qbezier(-59,10)(-62,7)(-65,10)\qbezier(-65,10)(-68,13)(-71,10)
\qbezier(-71,10)(-75,7)(-78,10)
\linethickness{2pt}
\qbezier(-40,20)(-50,2)(-75,1)
\qbezier(-15,10)(-18,55)(-40,20)
\qbezier(-15,10)(-12,-30)(0,-30)
\qbezier(15,10)(12,-30)(0,-30)
\qbezier(15,10)(18,55)(40,20)
\qbezier(40,20)(50,2)(75,1)
\end{picture}
\caption{\label{9}In anti-de Sitter space, the effective binding potential gets
modified, and would-be bound state with $m^2 \neq 0$
becomes quasi-localized. Particle
has finite probability to escape the brane via tunneling.}
\end{center}
\end{figure}

When gravity of the brane
is turned on, the situation changes. The action in the background
metric (\ref{RS}) is
\be
S_{\phi} = \int~d^4 x~dz~a^4\left[ \frac{1}{a^2} \eta^{\mu \nu}
\partial_{\mu} \phi \partial_{\nu} \phi - \frac{1}{2} (\partial_z
\phi)^2 - \frac{1}{2} V(z) \phi^2 \right]
\label{80*}
\ee
where, as before, $a(z) = \mbox{e}^{-k|z|}$. Since $a^{-2}$ grows at
large $z$, the first term in the integrand of eq.~(\ref{80*})
dominates away from the brane over the potential term, and continuum
of KK modes starts from zero. The eigenvalue equation
for four-dimensional masses reads now
\be
\frac{1}{a^4} \partial_z (a^4 \partial_z \phi)
- V(z) \phi + \frac{m^2}{a^2} \phi = 0
\label{81*}
\ee
It is useful to note that the normalization condition for the
eigenfunctions $\phi_m(z)$  is
\be
  \int~dz~a^2(z) \phi_m(z) \phi_{m'} (z) = \delta_{m m'}
\label{81a*}
\ee
Indeed, the eigenvalue equation (\ref{81*}) can be written in the form
\[
   - \partial_z (a^4 \partial_z \phi_m)
 + a^4 V(z) \phi_m = m^2 a^2 \phi_m
\]
The operator entering the left hand side is Hermitean, so the
eigenfunctions are orthogonal with the measure $a^2 dz$. This is
precisely
the same measure as multiplies the kinetic term in the action, i.e.,
the first term in the integrand of eq.~(\ref{80*}).

At large $z$, the second term in eq.~(\ref{81*})
is negligible as compared to the
third one.
Effectively, this means that the binding potential gets modified
and tends to zero as $|z| \to \infty$, see Fig. 9. 
 The wave functions at $z \to \infty$ are
\[
   \phi(z) = \mbox{const} \cdot \mbox{e}^{3kz/2} 
\sin \left(\frac{m}{k} \mbox{e}^{kz} + \varphi_m \right)
\]
(these are normalized to delta-function with weight $a^2 dz$).
The point is that
the continuum spectrum is determined by the large-$z$ asymptotics of
eq.~(\ref{81*}), and it starts from $m^2=0$ irrespectively of the form
of $V(z)$ (provided that $V(z)$ does not rapidly increase as $z \to
\infty$). As there are no bound states embedded in the continuum, the
massive bound states of eq.~(\ref{80+}) become now resonances, i.e.,
quasi-localized states having finite widths of decay (finite
probability of escape from the brane to $z = \infty$). These widths
depend on the potential $V(z)$ binding the particles to the brane.

It is worth noting that in RS2 set up, massless scalar field has a
zero mode even without the potential $V(z)$ \cite{Bajc}. Its wave
function is $\phi = \mbox{const}$, and it is normalizable with the
appropriate weight $a^2 dz$. Once this field is given a mass, the
would-be bound  state becomes metastable against escape into extra
dimension. For example, for constant $V(z) \equiv \mu^2$, the mass of
the four-dimensional particle is equal to \cite{DRT1}
\be
  m = \frac{\mu}{\sqrt{2}}
\label{83*}
\ee
and its width of escape to $z=\infty$ is
\[
  \Gamma = m\frac{\pi}{16}\left(\frac{m}{k}\right)^2
\]
The latter formula illustrates a general feature of these decays: for
small mass of the would-be bound state, the probability of its decay
into extra dimension is small, the reason being that the decay occurs
through tunneling, and hence it is naturally suppressed.

The above general arguments imply that the metastability of massive
particles against escape into extra dimension should be characteristic
to all kinds of matter, including fermions, provided these have bulk
modes. The calculation of the life-time of a fermion bound to the
brane by the mechanism of subsection 3.1, has been performed in
Ref.~\cite{DRT1}. This life-time depends not only on the fermion mass
and anti-de Sitter radius, but also on other parameters, so quantitative
estimates are premature at this stage.

Yet another interesting property of anti-de Sitter bulk has to do with
virtual KK states. With massive four-dimensional bosons only, the
potential between sources is of the Yukawa type.  Once there exist
arbitrarily light KK states, one expects the potential to have long
ranged tail. For example, in the scalar field model (\ref{80*}) with
constant $V(z) \equiv \mu^2$, one finds the potential between two
distant sources of the scalar field, $q_1$ and $q_2$,
located on the brane \cite{DRT1}
\be
V(r) = - \pi q_1 q_2 k \frac{\mbox{e}^{-mr}}{r}
 - 60 \pi q_1 q_2 \cdot \frac{1}{km^4} \cdot \frac{1}{r^7}
\label{85*}
\ee
where $m$ is given by eq.~(\ref{83*}). The first (Yukawa) term here is
due to the exchange by massive quasi-localized mode, whereas the
second one is due to the exchange by the KK continuum states. In a
model meant to describe a massive four-dimensional particle, the
potential has power-law behaviour at large distances!
The assumption that the field has bulk modes, which has been crucial
for the above discussion, is not, in fact, an innocent one, especially
if particles are charged. Indeed, if gauge fields are localized by the
mechanism of subsection 3.2, charged particles are confined in the
bulk, so they {\it do not} have bulk modes. In this scenario, escape of
a charged particle from the brane into extra dimension is
impossible. Later on we shall describe another mechanism of the gauge
field localization, which will allow for a charged particle to escape
to $z \to \infty$.

\subsection{Holographic interpretation}

Is it possible to describe the low energy physics on the brane
entirely in four-dimensional language? Certainly not, if one thinks in
terms of usual weakly coupled theories allowing for particle
interpretation: continuum KK modes do not correspond to particles
travelling in four-dimensional space-time along the brane. On the
basis of adS/CFT correspondence 
\cite{Maldacena:1998re,Gubser:1998bc,Witten:1998zw}
it has been argued \cite{ads,Gubser:1999vj}, however, that RS2 scenario
may be described by a strongly coupled four-dimensional conformal
field 
theory (CFT)
with an ultraviolet cut-off, interacting with conventional
gravitational field. The correction (\ref{75*}) to Newton's gravity
law is then interpreted as coming from ``one loop'' contribution of
conformal matter to the graviton propagator 
\cite{ads,Gubser:1999vj,Arkani-Hamed:2000ds}. Indeed, in conformal
field theory language,
this correction to
the graviton propagator
has the following form (indices are omitted),
\[
\delta G(x-y) = \mbox{const} 
\int~d^4u~d^4v~D(x-u) \langle T(u) T(v) \rangle D(v-y)
\]
where $D(x-y)$ is a free four-dimensional graviton propagator, and
$T(u)$ is the energy-momentum tensor of conformal fields. Hence,
\[
  \Box_x \Box_y (\delta G(x-y)) 
= \mbox{const} \cdot \langle T(x) T(y) \rangle
\]
Now, the correlator of the energy-momentum tensor of a conformal field 
theory is
 \[
\langle T(x) T(y) \rangle = \frac{\mbox{const}}{(x-y)^8}
\]
so that
\[
\delta G(x-y) = \frac{\mbox{const}}{(x-y)^4}
\]
The contribution to the gravitational potential is the time integral
of this expression, which immediately gives
\[
\Delta V(r) = \frac{\mbox{const}}{r^3}
\]
in agreement with eq.~(\ref{75*}). This is the most straightforward
check of the interpretation of KK gravitons in terms of
four-dimensional conformal field theory; there are a number of other 
checks.

Likewise, escape of a particle into extra dimension has a CFT
interpretation as a decay into conformal modes, now interacting with
matter fields, whereas the power-law correction to the Yukawa
potential, eq.~(\ref{85*}), is again due to the exchange by these 
modes.

Another way to see how effective conformal matter shows up in
four-dimensional theory is to consider gravitational field of a
massive point-like particle which sits on the brane until some moment
of time (say, $t=0$) and then leaves the brane and escapes into extra
dimension along the geodesic normal to the brane
\cite{Gregory:2000rh,Giddings:2000ay}. The four-dimensional
gravitational field induced by this particle is straightforward to
calculate in linearized five-dimensional theory. One finds that
outside the light cone, i.e., at $({\bf x}^2 - t^2) > 0$, the
four-dimensional gravitational field induced on the brane is still
described by the
linearized Schwarzschild metric (in other words, four-dimensional
gravitational field does not change outside the light cone, in accord
with causality). Inside the light cone, the induced four-dimensional
metric is flat. If one {\it defines} the effective four-dimensional 
energy-momentum tensor by
\[
8 \pi G_{(4)} T_{\mu \nu}^{(eff)} \equiv 
R_{\mu \nu}^{(4)} - \frac{1}{2} g^{(4)}_{\mu \nu} R^{(4)}
\]
where $g^{(4)}_{\mu \nu}$ is the four-dimensional metric induced on
the brane, then $ T_{\mu \nu}^{(eff)}$ corresponds to a thin shell
of matter expanding along the four-dimensional light cone and
dissipating as $1/r^2$. This is precisely the behaviour of
energy-momentum  
expected in any conformal field theory \cite{Coleman:1977yb}.

The holographic approach is useful for analysing aspects of
phenomenology of  RS2 model 
\cite{Giddings:2000mu,Giddings:2000ay,Arkani-Hamed:2000ds} and
sheds new light on  cosmology with infinite extra dimesnions (see, e.g.,
Refs.~\cite{Hawking:2000kj,Nojiri:2000gb,Anchordoqui:2000du} and
references therein).
Conformal field theory language is useful, to a certain extent, also
in RS1 scenario with two branes and compact extra dimension (see,
e.g.,
Ref.~\cite{Arkani-Hamed:2000ds} and references therein). In that case,
the four-dimensional interpretation is based on strongly coupled
theory with broken conformal invariance.

These examples show that four-dimesional conformal field theories,
weakly coupled to ordinary matter, may emerge naturally in various
multi-dimensional contexts. Irrespectively of extra dimensions, it is
of interest to understand phenomenological implications of possible
existence in nature of a conformal sector with broken or unbroken
conformal invariance.

\subsection{More than one extra dimensions. Localization of gauge
fields by gravity}

One approach to extend the construction outlined above to space-time of
more than one extra dimension is to consider intersecting branes
of codimension one
\cite{Arkani-Hamed:2000hk}. In $(3+n)$-dimensional space, each of
these branes has dimesnion $(3+n-1)$, so none of them by itself is a
candidate for our world. However, an intersection of $(n-1)$ of these
branes is a three-dimensional manifold, to which our matter may be
bound. Furthermore, gravity is naturally localized on this
intersection manifold, so this set up is phenomenologically viable
\cite{Arkani-Hamed:2000hk}.

Localizing gravity on a genuine three-brane 
embedded in space with more than one extra dimensions
is more difficult. For many solutions of the
Einstein equations in space-time of more than one transverse
dimensions, the gravitational effect of the brane vanishes at large
distances. To some extent, the situation here is similar to classical
solutions in electrodynamics: while the electric field of a charged
plane extends to infinity, the electric fields of a charged line and a
charged point decay as $1/r$ and $1/r^2$, respectively. Similarly, 
with more than one extra dimensions,
the
gravitational field of a massive brane often decreases towards
infinity in transverse space, so there is no reason for the
localization of a graviton on the brane.

One interesting exception is the geometry with compact, but warped,
additional dimension(s). In the case of two extra dimensions, metric
of such a set up, away from the brane is
\be
  ds^2 = a^2(z) [\eta_{\mu \nu} dx^{\mu} dx^{\nu} - d\theta^2] 
-dz^2
\label{91*}
\ee
where $z$ is the coordinate along a single non-compact extra
dimension, $\theta$ is a compact extra coordinate running from $0$ to
$2\pi R$ and the warp factor is still
\[
  a(z) = \mbox{e}^{-k|z|}
\]
This metric is an symptotics of
a solution to the (4+1+1)-dimensional Einstein
equations with negative bulk cosmological constant and a three-brane
with appropriately tuned energy-momentum tensor
\cite{Gherghetta:2000qi} (see Refs.~\cite{Cohen:1999ia,Gregory:2000gv} for related solutions). 
The brane itself is a sort of cosmic
string. The generalizations of this construction to more than two
extra dimensions have been found in 
Ref.~\cite{Gherghetta:2000jf,Randjbar-Daemi:2000ft}.

Another way to obtain metrics of the type (\ref{91*}) is to consider a
$(3+n)$-brane embedded in $(3+n+1)$-dimensional space \cite{DRT2}.
In other words, one considers a brane of codimension one, as in RS2
set up, but now in $(5+n)$-dimensional space-time. Then the metric
\be
  ds^2 = a^2(z) [\eta_{\mu \nu} dx^{\mu} dx^{\nu} - 
\delta_{ij} d\theta^i d\theta^j ] 
-dz^2
\label{92*}
\ee
which is a straightforward generalization of eq.~(\ref{RS}), obeys the
full Einstein equations with essentially the same fine-tuning
conditions as in the original RS2 set up. One then takes $\theta^i$ to
be compact coordinates, $\theta^i \in (0, 2\pi R_i)$, so that these
dimensions are invisible at low energies in complete analogy to the
Kaluza--Klein picture.

In either case, there exists a graviton zero mode which is independent
of $\theta^i$ outside the brane and decreases at large $z$ as
$\mbox{e}^{-2k|z|}$. Gravity is localized on the brane in the same way
as in  RS2 case.

A novel feature of this set up is the
localization  of gauge fields by
gravity \cite{Oda:2000zc,DRT2}. 
With appropriate normalization, the action of the
gauge field in the background (\ref{92*}) is
\[
S_{gauge} = - \frac{1}{4} 
\int~\prod_i~\frac{d\theta^i}{2\pi R_i}~dz~d^4x~\sqrt{g}
g^{AC} g^{BD} F_{AB}
F_{CD}
\]
At low energies, $E \ll 1/R_i$, the relevant gauge field
configurations are independent of $\theta^i$. We shall be interested
in four-vector part of the gauge field, so we choose the gauge $A_z
=0$ and set $A_{\theta^i} = 0$. Then the linearized gauge field
equations are
\begin{eqnarray}
- \frac{1}{a^{2+n}} \partial_z (a^{2+n} \partial_z A_{\mu})
+ \frac{1}{a^2} \eta^{\lambda \nu} \partial_{\lambda} F_{\nu \mu} &=& 0
\label{93*} \\
\partial_z(\eta^{\mu \nu} \partial_{\mu} A_{\nu}) &=& 0
\label{93++}
\end{eqnarray}
In complete analogy to eq.~(\ref{81a*}), the eigenfunctions of
eq.~(\ref{93*}) are to be normalized with the measure
\be
  \int~dz~a^n
\label{93**}
\ee
For $n>0$ (one or more compact warped dimensions), there exists a zero
mode, which is independent of $z$,
\[
  A_{\mu} = A_{\mu} (x)
\]
This mode is normalizable with the measure (\ref{93**}) and
corresponds to massless vector boson localized on the brane.

The fact that the zero mode of the gauge field is {\it constant} over
extra dimensions, ensures charge universality, which otherwise would
be an obstacle for the gauge boson localization, as discussed in
subsection 3.2. With constant gauge boson zero mode, the overlap
integrals like eq.~(\ref{18*}) become the norms of  matter zero modes,
whose values do not depend on the shapes of the zero modes. This way
to get around the charge universality problem relies, of
course, entirely on special geometry in the bulk.

Besides the zero mode, gauge fields in the background (\ref{92*}) have
arbitrarily light bulk modes; the gauge theory is not in
confinement phase in the bulk. This is in sharp contrast to the
mechanism discussed in subsection 3.2. Furthermore, the gravitational
mechanism of the gauge field localization allows the electric charge,
and other gauge charges, to leak into extra dimensions. This would mean
effective non-conservation of gauge charges in our world
\cite{DRT2}. A process 
``$e^- \to \mbox{nothing}$'' would be  a clear signature of infinite extra
dimensions.

A problem 
\cite{Ponton:2001gi}
with the set up with warped compact dimension(s) is that the
metric (\ref{92*}) has a singularity at $z= \infty$. 
Indeed, the sizes of compact dimensions, $\mbox{e}^{-k|z|} R_i$,
tend to zero as $z \to \infty$. Although the proper distance to the
singularity along the $z$-axis is infinite, proper distance along
time-like geodesics 
is finite. This implies that physics near the singularity may affect
physics on the brane. Possible resolutions of this singularity and
their effects on physics on the brane have been discussed recently in
Ref~\cite{Ponton:2001gi}. 

\section{Further developments}

Both versions of the brane world picture --- with
large and flat extra dimensions, and with warped extra dimensions ---
have 
been  developed along many different lines. Without pretending to
give an account of any degree of completeness,  
let us briefly discuss a few of them.

\subsection{Cosmological constant}

In the brane world context, the cosmological constant problem may be
reformulated as a problem of why the vacuum energy density has
(almost) no effect on the curvature induced on our brane. This
reformulation sounds suggestive, as it implies that the vacuum energy
density {\it may} affect the overall geometry of multi-dimensional
space-time, but this may occur in such a way that metric induced on
our brane is (almost) flat. Off hand, it seems plausible that, in the
case of non-factorizable geometry, the vacuum energy density may
induce a non-trivial warp factor, while the four-dimensional Poincar\'e
invariance remains intact. The latter possibility may exist
irrespectively of the brane world picture \cite{RubSh2,Randjbar-Daemi:1986wg}.

Recent attempts \cite{Arkani-Hamed:2000eg,Kachru:2000xs} to solve the
cosmological constant problem in the framework of brane world and
warped extra dimension, combine these ideas with the suggestion, put
forward in a similar context  in four-dimensional theories
\cite{Dolgov:1982gh,Peccei,Barr}, 
that
 a hypothetic scalar field
conformally coupled to our matter may play an important role. 
This field is taken to be a bulk
field, so the action is a sum of the bulk and brane contributions,
\[
  S = S_{bulk} + S_{brane}
\]
Here the bulk term is
\[
S_{bulk} = \int~dz~d^4x~\sqrt{g^{(5)}}
\left( - \frac{1}{2\kappa_{(5)}} R^{(5)} + 
\frac{b}{2} g^{AB} \partial_A \phi \partial_B \phi \right)
\]
where $\kappa_{(5)} = 8 \pi G_{(5)}$, $b$ is an arbitrary constant and
space-time is five-dimensional. The brane contribution is assumed to
have the following structure,
\[
S_{brane} =  \int~d^4x~\sqrt{g^{(4)} \mbox{e}^{4\kappa_{(5)}
\phi(0)}}
{\cal L} (\psi, g_{\mu \nu}^{(4)} \mbox{e}^{\kappa_{(5)} \phi(0)})
\]
where ${\cal L}$ is the complete Lagrangian of matter
on the brane, $\psi$ stands
for all fields on the brane, $g^{(4)}_{\mu \nu}$ is the induced metric
and $\phi(0) \equiv \phi (x, z=0)$ is the induced scalar field. In the
absence of real particles on the brane, the brane Lagrangian reduces
to the vacuum energy, with all quantum effects included,
 interacting with gravitational and scalar fields
\[
S_{brane} =  - \epsilon_{vac}~
\int~d^4x~\sqrt{g^{(4)} \mbox{e}^{4\kappa_{(5)}
\phi(0)}}
\]
It is clear now, that the magnitude of $\epsilon_{vac}$ is
unimportant, as it can be compensated by a shift of $\phi$: if for
some $\epsilon_{vac}$
there
exists a solution with four-dimensional Poincar\'e invariance, then such
a solution exists for any $\epsilon_{vac}$ of the same sign.

There indeed exist solutions to the resulting system of equations,
which have four-dimensional Poincar\'e invariance
\cite{Arkani-Hamed:2000eg,Kachru:2000xs}. In particular, for $b=3$,
these are the only solutions with maximum four-dimensional symmetry
\cite{Arkani-Hamed:2000eg}: solutions with four-dimensional de Sitter
or anti-de Sitter symmetry are absent. For $b=3$, the explicit form of
metric is
\be
  ds^2 = \hat{a}^2 (z) \eta_{\mu \nu} dx^{\mu} dx^{\nu} - dz^2
\label{100*}
\ee
where the warp factor is
\[
\hat{a} (z) = \left( 1 - \frac{2}{3\kappa_{(5)}^{2/3}}
\mbox{e}^{2 \kappa_{(5)} \tilde{\phi}_0} \cdot |z| \right)^{1/4}
\]
and $\tilde{\phi}_0 $ is an arbitrary parameter of solution related to
the value of $\phi(z=0)$ as follows,
\[
  \tilde{\phi}_0 = \phi(0) + \frac{1}{2 \kappa_{(5)}}
\ln (\epsilon_{vac} \kappa_{(5)}^{8/3})
\]
Induced metric on the brane is flat irrespectively of the value of
$\epsilon_{vac}$.

One problem with this mechanism is that metric (\ref{100*}) has a
naked singularity at finite proper distance to the brane,
\[
   |z_{sing}| = \frac{3}{2} \kappa_{(5)}^{2/3} \mbox{e}^{-2 \kappa_{(5)}}
\]
It has been argued that a possible resolution of this singularity
reintroduces the cosmological constant problem 
\cite{deAlwis,Forste:2000ps,Csaki:2000wz}. As an example, if one
introduces another brane placed at $|z| < |z_{sing}|$ (and imposes the
orbifold symmetry), the singularity will be absent between the branes,
i.e., in the entire orbifold. However, the tension of the second brane
has to be fine-tuned for the metric (\ref{100*}) to remain the
solution of the complete set of the Einstein equations. This
fine-tuning is no better than the usual
fine-tuning of the cosmological constant in four-dimesnional theories.

Another potential difficulty, common to many attempts to solve the
cosmological constant problem, is that in the presence of matter, the
energy density (in our context, energy density on the brane) contains 
the vacuum contribution and the contribution of real
matter. It is hard to find any distinction between the two; in other
words, the same mechanism that compensates for the cosmological
constant will tend to modify the gravitational interactions of
ordinary matter.
This causes doubts that the gravitational interactions in these
models reduce to four-dimensional General Relativity at large
distances, and, in particular, that any compensation
 mechanism will be cosmologically
viable (see, however, Ref.~\cite{Rubakov:2000aq}).

\subsection{Gravity at ultra-large distances}

In relatively simple models discussed in Sections 5 and 6, gravity on
the brane is effectively four-dimensional at large enough distances
due to the presence of the graviton zero mode. One may wonder whether
in other cases the graviton spectrum may be more complicated, and, in
particular, whether four-dimensional gravity may be modified at
ultra-large scales.

Original (and failed, see below) proposals
\cite{CGR,Kogan:2000wc,GRS}, exploring this possibility, invoked
dynamical negative tension branes (and one extra dimension). One of
them \cite{Kogan:2000wc} was to consider two positive-tension branes
of equal tensions $\sigma$ and a negative-tension brane of tension
($-\sigma$) in between the two. For simplicity, the positive-tension
branes are placed at fixed points of an orbifold; in any case, extra
dimension is compact. With negative bulk cosmological constant tuned
according to eq.~(\ref{53*}), the warp factor has a minimum at the
position $z_{-}$ of the negative-tension brane,
\[
  a(z) = \mbox{e}^{k|z-z_{-}|} \; , \;\;\;\;\; 
0 \leq z \leq z_c
\]
where $z=0$ and $z=z_c$ are positions of the two positive-tension
branes. In the limit $z_c \to \infty$,
$z_{-} \to \infty$, $(z_c - z_{-}) \to \infty$, all the three branes
are infinitely far apart, and
there  exist two graviton zero modes concentrating near $z=0$ and
$z=z_c$, respectively. At finite, but fairly large
$z_{-}$ and $z_c$, namely $z_{c},z_{-}, (z_c - z_{-}) \gg k^{-1}$,
the degeneracy between
the graviton modes is lifted, as is usually the case
 in quantum mechanics. 
One linear combination of the modes remains massless, whereas another
linear combination acquires a mass $m_g$ which is much smaller than 
the masses of KK excitations, $m_{KK}$. Hence, at intermediate
distances,
$m_{KK}^{-1} \ll r \ll m_g^{-1}$, both the lightest and massless gravitons
contribute  into the four-dimensional Newton's law, while at
ultra-large distances, $r \gg m_g^{-1}$, only massless graviton does
(in fact, the situation is more subtle, as there exists a ''ghost''
massless scalar mode that also 
contributes into interactions 
at ultra-large distances, see below). The four-dimensional gravity
gets modified at distance $m_g^{-1}$, which may be naturally of order
than 100 Mpc or larger.

Another set up \cite{CGR,GRS} contains a brane of positive tension
$\sigma$ and another brane of negative tension $(-\sigma/2)$. The bulk
cosmological constant between the branes is again tuned as in
eq.~(\ref{53*}), but $\Lambda$
is set equal to zero outside the negaive-tension
brane. Then the four-dimensionally flat solution has a
non-trivial warp factor
between the branes,
\[
  a(z) = \mbox{e}^{-k|z|} \; , \;\;\;\;\; 0 \leq z \leq z_{-}
\]
whereas the five-dimensional space-time is flat outside the
negative-tension brane,
\[
  a(z) = \mbox{const}  \; , \;\;\;\;\; z > z_{-}
\]
This set up has infinite extra dimension.
 
In the limit $z_{-} \to \infty$, this set up reduces to RS2, and there
exists the graviton zero mode near the positive-tension brane. For
finite $z_{-}$, this would-be zero mode becomes a resonance of small
but finite width $\Gamma_g$ \cite{GRS,Csaki:2000pp,Dvali:2000rv},
i.e., the four-dimensional graviton becomes metastable against 
escape into extra dimension. At intermediate distances, 
$k^{-1} \ll r \ll \Gamma_g^{-1}$, gravity on the positive-tension
brane is four-dimensional. At ultra-large distances, 
$r \gg \Gamma_g^{-1}$, 
Newton's gravity law is no longer valid; in fact,
attraction between masses changes into repulsion at 
$r \sim \Gamma_g^{-1}$ \cite{Gregory:2000iu,Csaki:2001cx}.

The problem 
\cite{Witten:2000zk}
with these two scenarios (as well as with intermediate
ones \cite{Kogan:2000cv}) is the existence of dynamical
(``vibrating'') negative-tension brane(s). It has been shown
explicitly \cite{Pilo}, in the context of the second set up, that 
among the excitations about the classical solution,
there exists
 a four-dimensional scalar field with {\it negative} kinetic term,
i.e., that energy is not positive-definite and, furthermore,
is not bounded from below. At quantum
level, spontaneous creation of negative energy quanta is possible, so
the whole set up is unstable. 

Another way \cite{Dvali:2000rv,Dvali:2000km} to see that modification
of gravity at ultra-large distances is difficult, is  to recall the 
van~Dam--Veltman--Zakharov phenomenon \cite{vanDam:1970vg,Zakharov}.
Namely, effective four-dimensional description of gravity with
non-trivial properties at ultra-large distances would involve tensor
field(s) with small but non-vanishing mass (or width). However, the
propagator of massive tensor field in flat background does not
coincide with the massless propagator even for $p \gg m$: the
(relevant part of the) former is
\[ 
   G_{\mu\nu\lambda\rho}^{(m)}(p \gg m)
=\frac{1}{p^2} \left[ \frac{1}{2} (\eta_{\mu \lambda} \eta_{\nu \rho}
+ \eta_{\mu \rho} \eta_{\nu \lambda}) - 
\frac{1}{3} \eta_{\mu \nu} \eta_{\lambda \rho} \right]
\]
whereas the latter has another tensor structure,
\[ 
   G_{\mu\nu\lambda\rho}^{(0)}
=\frac{1}{p^2} \left[ \frac{1}{2} (\eta_{\mu \lambda} \eta_{\nu \rho}
+ \eta_{\mu \rho} \eta_{\nu \lambda}) - 
\frac{1}{2} \eta{\mu \nu} \eta_{\lambda \rho} \right]
\]
Hence, even at intermediate distances, a theory with massive graviton
differs from General Relativity; in particular, the former would lead
to wrong (and phenomenologically unacceptable) prediction for bending
of light 
 \cite{Dvali:2000rv,Dvali:2000km}. In flat background, the correct
tensor structure may be restored by adding a ``ghost'' scalar field
with negative kinetic term, which interacts with the trace of
energy-momentum tensor  with appropriate strength $(1/6)$. This is
precisely what happens in the above scenarios.

A way out is to consider curved four-dimensional backgrounds. As an
example, a theory with massive graviton in four-dimensional anti-de
Sitter space-time has correct tensor structure of the propagator,
provided the graviton mass is smaller than the inverse adS radius
\cite{Kogan:2001uy,Porrati:2001cp} (see, however,
Ref.~\cite{Dilkes:2001av}). In the context of extra dimensions, curved
four-dimensional space-time emerges when the condition (\ref{53*}) is
relaxed. Indeed, it has been found in
Refs.~\cite{Kogan:2001vb,Karch:2000ct} that even in models with
positive-tension branes only, relaxing the condition  (\ref{53*})
leads to four-dimensional gravity which differs from General
Relativity at ultra-large scales (``behind the horizon''). It remains
to be understood whether this property has observable and/or
cosmologically significant consequences.

\subsection{Approximate Lorentz invariance}

Extra dimensions provide, among other things, a
framework for studying whether small violation of Lorentz invariance
is possible, and what phenomenological consequences it may have. It
has been understood long ago \cite{viss} that the geometry of the
multi-dimensional space-time may not have four-dimensional Lorentz
invariance, and yet the four-dimensional geometry induced on ``our''
brane may be invariant under  Lorentz symmetry. 

As an example, in five
dimensions, the warp factors of time and three-dimensional space may
be different, so the metric may have the form
\be
  ds^2 = a^2(z) dt^2  - b^2 (z) d{\bf x}^2 - dz^2
\label{102*}
\ee
Classical solutions with $a(z) \neq b (z)$ have been found in 
Refs.~\cite{viss,Bowcock:2000cq,Chung:2000ji,Csaki:2000dm,Dubovsky:2001fj};
these involve additional matter in the bulk. The four-dimesnional
coordinates may be chosen in such a way that at the brane position,
$z=0$, the warp factors are
\[
   a(0) = b(0) =1
\]
and the four-dimensional metric induced on the brane is
Minkowskian. If the wave functions of ``our'' particles have zero
width along the fifth direction, it is this Minkowski metric that
governs the dynamics of these particles, so the effective
four-dimensional theory is exactly Lorentz-invariant. If, however, the
wave functions have finite extent in extra dimension, deviations of
$a(z)$ and $b(z)$ from 1 are felt by  particles on the brane, and
four-dimensional Lorentz invariance is only approximate. This, in
particular, applies to gravitons: if $a(z)$ (not necessarily $b(z)$)
 decreases sufficiently rapidly towards $z \to \infty$, gravitons are
bound to the brane by essentially the same mechanism as in RS2 set up,
and the wave function of the graviton zero mode has finite spread in
the fifth dimension.

One Lorentz-violating effect that occurs at relatively 
low spatial momenta is that the dispersion relation gets modified:
instead of the usual $\omega^2 = m^2 + {\bf p}^2$ (where $\omega$ and
${\bf p}$ are energy and three-momentum of a particle on the brane),
the dispersion relation is now
\be
    \omega^2 = m^2 + c^2 {\bf p}^2
\label{104*}
\ee
where the parameter $c$ depends on the spread of the wave function in
the fifth dimension \cite{Chung:2000ji,Csaki:2000dm}.
The form of the dispersion
relation (\ref{104*}) is in accord with the general analysis of 
Ref.~\cite{Coleman:1999ti}. To see how eq.~(\ref{104*}) appears, let
us again consider a prototype model of a scalar field with action
(\ref{p28*}), but now in the background metric (\ref{102*}).
The equation for the wave function $\phi$ ant  energy $\omega$ is
 now
\be
  \frac{b^3 (z)}{a(z)}~\omega^2 \phi = a(z)~b(z)~{\bf p}^2 \phi +
  {\cal H}(z) \phi
\label{104**}
\ee
where
\[
{\cal H} = - \partial_z (a b^3 \partial_z) + ab^3 V(z)
\]
Note that ${\cal H}$ is Hermitean. Let us assume that at ${\bf p} =0$,
equation (\ref{104**}) has a discrete eigenvalue $\omega^2 = m^2$; the
corresponding wave function $\phi_m (z)$ concentrates near the
brane. At small but non-vanishing ${\bf p}$, one considers the first
term on the right hand side of eq.~(\ref{104**}) as perturbation, and
obtains the lowest order correction to the energy,
\[
\Delta \omega^2 = c^2 {\bf p}^2
\]
where
\[
   c^2 = \frac{\int~dz~ a b |\phi_m(z)|^2}{\int~dz~ a^{-1} b^3 |\phi_m(z)|^2} 
\]
For $a(z) \neq b(z)$ , the parameter $c$ is non-universal, as it
depends on the shape of the wave function $\phi_m$. For narrow wave
functions, one has
\[
  c^2 = 1 + 2[a'(0) - b'(0)] 
\frac{\int~dz ~z |\phi_m(z)|^2}{\int~dz |\phi_m(z)|^2}
\]
The violation of Lorentz invariance is small provided the correction to
the relation $c^2 = 1$ is small\footnote{Another possibility is that
the wave functions of all particles bound to the brane  are very
similar.}. 
The graviton wave function is not,
however, narrow in this set up, so the violation of Lorentz invariance
in gravitational sector may be substantial. Note that the modification
of the dispersion law may take place 
for both  massive and massless particles.

Another effect \cite{Dubovsky:2001fj} occurs if $a(z)$ and $b(z)$ 
have different asymptotics as $|z| \to \infty$. Suppose that $a(z)$
decreases  faster than $b(z)$, i.e.,
\[
   a(z) \to 0 \; , \;\;
\frac{a(z)}{b(z)} \to 0 \;\;\; \mbox{as} \;\; |z| \to \infty
\]
Suppose further that $V(z)$ tends to a constant as $|z| \to \infty$.
Then, in complete analogy to subsection 6.2, the continuum spectrum of
eigenvalues $\omega^2$ of eq.~(\ref{104**}) starts from zero
(the term proportional to $V(z)$ is irrelevant at large $|z|$ since
$ab^3 V \ll (b^3/a) \omega^2$ at $|z| \to \infty$ for arbitrarily
small $\omega^2$). For ${\bf p}^2 = 0$, there may exist a localized
zero mode which would describe a massless particle residing on a
brane. Unlike in subsection 6.2, however, the continuum starts from
zero {\it also for} ${\bf p}^2 \neq 0$: the term involving ${\bf p}^2$
in eq.~(\ref{104**}) is also negligible as compared to  
$(b^3/a) \omega^2$ at large $|z|$. On the other hand, the zero mode
gets lifted: real part of its energy is equal to
$\omega = c |{\bf p}|$. 
Hence, this mode becomes {\it quasi}-localized for 
${\bf p}^2 \neq 0$ (there are no bound states embedded in continuum). 
This means that even massless particles are metastable in this set up
against escape into extra dimension, provided the corresponding fields
have bulk modes. The larger the three-momentum $|{\bf p}|$, the larger
the width $\Gamma (|{\bf p}|)$ \cite{Dubovsky:2001fj}. This mechanism
of metastability of moving particles in theories with
broken four-dimensional Lorentz invariance may have interesting
phenomenological consequences, especially for physics of ultra-high
energy cosmic rays.

\subsection{Creation of a brane Universe}

At the late stage of the evolution of the Universe,
corresponding to temperatures well below the maximum temperature
$T_{*}$ in ADD scenario or well below TeV in RS1 and RS2 models,
brane cosmology is governed by the same four-dimensional laws as the
standard FRW cosmology \cite{Csaki:1999jh,Cline:1999ts,Csaki:2000mp}
(in the case of RS1, this assertion holds only if the radion is
stabilized, otherwise the late cosmology is very unconventional 
\cite{Binetruy:2000ut,Chung:2000zs}). An important assumption here is that the bulk
is empty, and, in the case of infinite extra dimensions, that the
horizon ``does not shine'' on our brane. Are these assumptions
reasonable? Is it conceivable that our world with all its reach
structure is merely a brane embedded in empty multi-dimensional space?

This brings us to a problem of the beginning of the brane Universe.
The requirement
 that the bulk should be 
essentially empty suggests a picture of
spontaneous creation of ``our'' brane in a multi-dimensional space
which initially has neither branes nor other excitations 
\cite{Gorsky:2000rz}. A model in which this possibility is realized is
a five-dimensional theory of four-form field (four-index
anti-symmetric tensor field) $B$ and branes, both charged and neutral
with respect to this field. Just like a charged particle couples to a
vector field through its world-line, 
$S_{int} = e \int~A_{\mu} dx^{\mu}$, a charged brane couples to
the four-form field 
 through its four-dimensional world-volume,
\[
   S_{int} = e \int~B_{ABCD} d \Omega^{ABCD}
\]
The picture is then as follows. Suppose that an initial state of a
five-dimensional Universe contains the $B$-field of  constant field
strength, 
\[
  H_{ABCDE} \equiv \partial_{[A} B_{BCDE]} = 
h \sqrt{g^{(5)}} \epsilon_{ABCDE}
\]
If it were not for the charged branes, the field $B$ would act as a
five-dimensional cosmological constant, as its field equations would
require that
\[
         h = \mbox{const}
\]

\begin{figure}[!hb]
\begin{center}
\unitlength1mm
\begin{picture}(60,60)(-25,-25)
\put(-10,-5){Charged}
\put(-7,-10){brane}
\put(15,25){Charged}
\put(18,20){brane}
\put(10,10){RS}
\put(8,5){brane}
\put(-1,20){\line(1,1){5}}
\put(0,15){\line(1,1){9}}
\put(1,10){\line(1,1){12}}
\put(3,6){\line(1,1){14}}
\put(6,3){\line(1,1){14}}
\put(10,1){\line(1,1){12}}
\put(15,0){\line(1,1){9}}
\put(20,-1){\line(1,1){5}}
\qbezier(-18.3,-18.3)(-10,-22)(0,-22)
\qbezier(-18.3,-18.3)(-12,-20)(-5,-18)
\qbezier(-18.3,-18.3)(-13,-17)(-10,-15)
\qbezier(-18.3,-18.3)(-22,-10)(-22,0)
\qbezier(-18.3,-18.3)(-20,-12)(-18,-5)
\qbezier(-18.3,-18.3)(-17,-13)(-15,-10)

\qbezier(33.5,33.5)(36,29)(35,20)
\qbezier(33.5,33.5)(34,32)(32,29)
\qbezier(33.5,33.5)(29,36)(20,35)
\qbezier(33.5,33.5)(32,34)(29,32)

\linethickness{1pt}
\qbezier[40](0,25)(24,24)(25,0)
\qbezier(0,-25)(24,-24)(25,0)
\qbezier(0,25)(-24,24)(-25,0)
\qbezier(0,-25)(-24,-24)(-25,0)
\qbezier(19,-1)(0,0)(-1,19)
\qbezier(19,-1)(38,0)(39,19)
\qbezier(19,39)(0,38)(-1,19)
\qbezier(19,39)(38,38)(39,19)
\end{picture}
\caption{\label{10} Spontaneously created system of
two charged branes with neutral (RS)
brane between them.}
\end{center}
\end{figure}
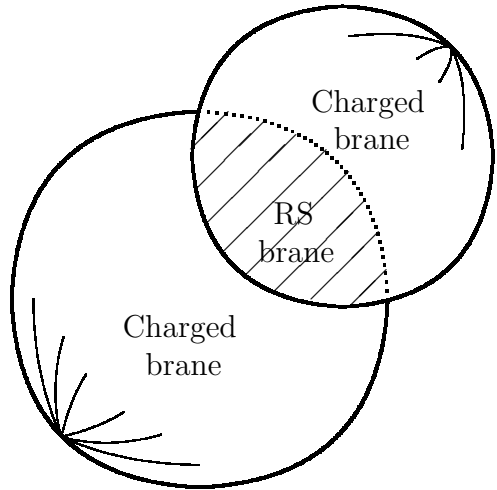

For non-vanishing $h$, charged branes are created spontaneously
\cite{Brown:1988kg}, much in the same way as $e^{+} e^{-}$-pairs are
spontaneously created by background electric field. Furthermore, a
neutral brane has a chance to be created in between the charged ones
\cite{Gorsky:2000rz}; it is this neutral brane  that is interpreted as
our Universe.
The system of the three branes at the moment of their materialization
is shown in Fig. 10.
After spontaneous creation, the three-volumes of all the
branes increase, the junction between the neutral and charged branes
moves away, and most of the neutral brane becomes a homogeneous
three-dimensional space. The bulk surrounding it remains empty.

If $h$ is not large, the probability of brane creation is
exponentially small. So, it is natural that other branes, which may be
created in the five-dimensional space, are extremely far away from
``ours''. Also, with sufficiently complicated theory on ``our'' brane,
it is conceivable that its initial state is such that a period of the
usual four-dimensional inflation on this brane occurs after it is
created. Hence, the model of Ref.~\cite{Gorsky:2000rz} is capable of
reproducing the conventional cosmological picture, with our Universe
being a brane embedded in empty bulk. This model  certainly represents
an
interesting twist in the discussion of the origin of our Universe.

\section{Conclusion}

Theories with large and infinite extra dimensions look rather exotic,
at least for the moment. Yet they lead to important insights on what
unusual phenomena may occur both at the energy scales accessible to
future accelerators, and at low energies, the domain of possible rare
effects. They also provide a framework for addressing a number of
phenomenological issues, like consistency of apparent 
non-coneservation
of energy and electric charge, naturalness of small violation of
Lorentz invariance and possible manifestations of a strongly coupled
conformal sector which interacts weakly with the Standard Model
fields. Furthermore, new ideas emerge in approaching fundamental
problems, such as the cosmological constant problem or the beginning
of our Universe. 

All this makes the subject interesting and lively. The question is
whether Nature follows any of the routes being explored in this 
context.

The author is indebted to L.~B.~Okun for his  suggestion to
write this mini-review. He thanks numerous colleagues at INR and
elsewhere for illuminating discussions. This work
has been supported in part by the Council for
Presidential Grants and State Support of Leading Scientific Schools,
grant 00-15-96626, RFBR grant 99-01-18410 and Swiss Science Foundation
Grant 7SUPJ062239.


\begin{thebibliography}{99}
\bibitem{Volobuev:1986wc}
I.~P.~Volobuev and Y.~A.~Kubyshin,
Theor.\ Math.\ Phys.\ {\bf 68} (1986) 788;
Theor.\ Math.\ Phys.\ {\bf 68} (1986) 885;
JETP Lett.\ {\bf 45} (1987) 581.
\bibitem{Antoniadis:1990ew}
I.~Antoniadis,
Phys.\ Lett.\ B {\bf 246} (1990) 377.
\bibitem{Lykken:1996fj}
J.~D.~Lykken,
Phys.\ Rev.\ D {\bf 54} (1996) 3693
[hep-th/9603133].


\bibitem{Polchinski:1996na}
J.~Polchinski,
``TASI lectures on D-branes,''
hep-th/9611050.
\bibitem{Horava:1996qa}
P.~Horava and E.~Witten,
Nucl.\ Phys.\ B {\bf 460} (1996) 506
[hep-th/9510209].
\bibitem{Lukas:1999yy}
A.~Lukas, B.~A.~Ovrut, K.~S.~Stelle and D.~Waldram,
Phys.\ Rev.\ D {\bf 59} (1999) 086001
[hep-th/9803235].
\bibitem{Visser:1985qm}
M.~Visser,
``An Exotic Class Of Kaluza-Klein Models,''
[hep-th/9910093].
\bibitem{Rubakov:1983bb}
V.~A.~Rubakov and M.~E.~Shaposhnikov,
Phys.\ Lett.\ B {\bf 125} (1983) 136.

\bibitem{Akama:1982jy}
K.~Akama,
Lect.\ Notes Phys.\ {\bf 176} (1982) 267
[hep-th/0001113].

\bibitem{Jackiw:1976fn}
R.~Jackiw and C.~Rebbi,
Phys.\ Rev.\ D {\bf 13} (1976) 3398.
\bibitem{Jackiw:1981ee}
R.~Jackiw and P.~Rossi,
Nucl.\ Phys.\ B {\bf 190} (1981) 681.
\bibitem{'tHooft:1976fv}
G.~'t Hooft,
Phys.\ Rev.\ D {\bf 14} (1976) 3432.

\bibitem{Libanov:2001uf}
M.~V.~Libanov and S.~V.~Troitsky,
Nucl.\ Phys.\ B {\bf 599} (2001) 319
[hep-ph/0011095].


\bibitem{Frere:2000dc}
J.~M.~Frere, M.~V.~Libanov and S.~V.~Troitsky,
``Three generations on a local vortex in extra dimensions,''
hep-ph/0012306.
\bibitem{Dvali:1997xe}
G.~Dvali and M.~Shifman,
Phys.\ Lett.\ B {\bf 396} (1997) 64
[hep-th/9612128].
\bibitem{Dvali:2000hr}
G.~Dvali, G.~Gabadadze and M.~Porrati,
Phys.\ Lett.\ B {\bf 485} (2000) 208
[hep-th/0005016].
\bibitem{Dvali:2001xg}
G.~Dvali and G.~Gabadadze,
Phys.\ Rev.\ D {\bf 63} (2001) 065007
[hep-th/0008054].
\bibitem{Dvali:2001rx}
G.~Dvali, G.~Gabadadze and M.~Shifman,
Phys.\ Lett.\ B {\bf 497} (2001) 271
[hep-th/0010071].
\bibitem{Dubrub} S.~L.~Dubovsky and V.~A.~Rubakov, to appear
\bibitem{Quevedo} F.~Quevedo and C.~A.~Trugenberger,
Nucl.\ Phys.\ B {\bf 501} (1997) 143
[hep-th/9604196]; \\
.~M.~Polyakov,
Nucl.\ Phys.\ B {\bf 486} (1997) 23
[hep-th/9607049].

\bibitem{Arkani-Hamed:1998rs}
N.~Arkani-Hamed, S.~Dimopoulos and G.~Dvali,
Phys.\ Lett.\ B {\bf 429} (1998) 263
[hep-ph/9803315].
\bibitem{Antoniadis:1998ig}
I.~Antoniadis, N.~Arkani-Hamed, S.~Dimopoulos and G.~Dvali,
Phys.\ Lett.\ B {\bf 436} (1998) 257
[hep-ph/9804398].
\bibitem{Hoyle:2001cv}
C.~D.~Hoyle, U.~Schmidt, B.~R.~Heckel, E.~G.~Adelberger, 
J.~H.~Gundlach, D.~J.~Kapner and H.~E.~Swanson,
Phys.\ Rev.\ Lett.\ {\bf 86} (2001) 1418
[hep-ph/0011014].
\bibitem{Mitrofanov} V.~P.~Mitrofanov and O.~I.~Ponomareva,
JETP {\bf 67} (1988) 1963.
\bibitem{Su:1994gu}
Y.~Su, B.~R.~Heckel, E.~G.~Adelberger, J.~H.~Gundlach, M.~Harris, G.~L.~Smith and H.~E.~Swanson,
Phys.\ Rev.\ D {\bf 50} (1994) 3614.
\bibitem{Long:1999dk}
J.~C.~Long, H.~W.~Chan and J.~C.~Price,
Nucl.\ Phys.\ B {\bf 539} (1999) 23
[hep-ph/9805217].
\bibitem{Long:2000xa}
J.~C.~Long, A.~B.~Churnside and J.~C.~Price,
hep-ph/0009062.
\bibitem{braginsky} V.~B.~Braginsky, private communication.
\bibitem{Giudice:1999ck}
G.~F.~Giudice, R.~Rattazzi and J.~D.~Wells,
Nucl.\ Phys.\ B {\bf 544} (1999) 3
[hep-ph/9811291].
\bibitem{Giudice:2001av}
G.~F.~Giudice, R.~Rattazzi and J.~D.~Wells,
Nucl.\ Phys.\ B {\bf 595} (2001) 250
[hep-ph/0002178].

\bibitem{Nussinov:1999jt}
S.~Nussinov and R.~Shrock,
Phys.\ Rev.\ D {\bf 59} (1999) 105002
[hep-ph/9811323].

\bibitem{Hewett:1999sn}
J.~L.~Hewett,
Phys.\ Rev.\ Lett.\ {\bf 82}, 4765 (1999)
[hep-ph/9811356].

\bibitem{Han:1999sg}
T.~Han, J.~D.~Lykken and R.~Zhang,
Phys.\ Rev.\ D {\bf 59} (1999) 105006
[hep-ph/9811350].


\bibitem{Arkani-Hamed:1999nn}
N.~Arkani-Hamed, S.~Dimopoulos and G.~Dvali,
Phys.\ Rev.\ D {\bf 59} (1999) 086004
[hep-ph/9807344].

\bibitem{Kaloper:2000jb}
N.~Kaloper, J.~March-Russell, G.~D.~Starkman and M.~Trodden,
Phys.\ Rev.\ Lett.\  {\bf 85} (2000) 928
[hep-ph/0002001].

\bibitem{Kakushadze:1999wp}
Z.~Kakushadze and S.~H.~Tye,
Nucl.\ Phys.\ B {\bf 548} (1999) 180
[hep-th/9809147].

\bibitem{Shiu:1999iw}
G.~Shiu, R.~Shrock and S.~H.~Tye,
Phys.\ Lett.\ B {\bf 458} (1999) 274
[hep-ph/9904262].

\bibitem{Cullen:2000ef}
S.~Cullen, M.~Perelstein and M.~E.~Peskin,
Phys.\ Rev.\ D {\bf 62} (2000) 055012
[hep-ph/0001166].


\bibitem{Giudice:2000ex}
G.~F.~Giudice, E.~W.~Kolb and A.~Riotto,
``Largest temperature of the radiation era and its cosmological  implications,''
hep-ph/0005123.
\bibitem{Cullen:1999hc}
S.~Cullen and M.~Perelstein,
Phys.\ Rev.\ Lett.\ {\bf 83} (1999) 268
[hep-ph/9903422].
\bibitem{Barger:1999jf}
V.~Barger, T.~Han, C.~Kao and R.~J.~Zhang,
Phys.\ Lett.\ B {\bf 461} (1999) 34
[hep-ph/9905474]
\bibitem{Hanhart:2001er}
C.~Hanhart, D.~R.~Phillips, S.~Reddy and M.~J.~Savage,
Nucl.\ Phys.\ B {\bf 595} (2001) 335
[nucl-th/0007016].






\bibitem{Hannestad:2001jv}
S.~Hannestad and G.~Raffelt,
``New Supernova Limit on Large Extra Dimensions,''
hep-ph/0103201.



\bibitem{Arkani-Hamed:1998vp}
N.~Arkani-Hamed, S.~Dimopoulos, G.~Dvali and J.~March-Russell,
``Neutrino masses from large extra dimensions,''
hep-ph/9811448.

\bibitem{Dienes:1999sb}
K.~R.~Dienes, E.~Dudas and T.~Gherghetta,
Nucl.\ Phys.\ B {\bf 557} (1999) 25
[hep-ph/9811428].


\bibitem{Dvali:1999cn}
G.~Dvali and A.~Y.~Smirnov,
Nucl.\ Phys.\ B {\bf 563} (1999) 63
[hep-ph/9904211].

\bibitem{Mohapatra:1999zd}
R.~N.~Mohapatra, S.~Nandi and A.~Perez-Lorenzana,
Phys.\ Lett.\ B {\bf 466} (1999) 115
[hep-ph/9907520]; \\
R.~N.~Mohapatra and A.~Perez-Lorenzana,
Nucl.\ Phys.\ B {\bf 576} (2000) 466
[hep-ph/9910474].

\bibitem{Barbieri:2000mg}
R.~Barbieri, P.~Creminelli and A.~Strumia,
Nucl.\ Phys.\ B {\bf 585} (2000) 28
[hep-ph/0002199].

\bibitem{Lukas:2000wn}
A.~Lukas, P.~Ramond, A.~Romanino and G.~G.~Ross,
Phys.\ Lett.\ B {\bf 495} (2000) 136
[hep-ph/0008049].

\bibitem{Cosme:2000ib}
N.~Cosme, J.~M.~Frere, Y.~Gouverneur, 
F.~S.~Ling, D.~Monderen and V.~Van Elewyck,
``Neutrino suppression and extra dimensions: A minimal model,''
hep-ph/0010192.

\bibitem{Perez-Lorenzana:2000hf}
A.~Perez-Lorenzana,
``Theories in more than four dimensions,''
hep-ph/0008333.

\bibitem{DDG}
K.~R.~Dienes, E.~Dudas and T.~Gherghetta,
Phys.\ Lett.\ B {\bf 436} (1998) 55
[hep-ph/9803466];
Nucl.\ Phys.\ B {\bf 537} (1999) 47
[hep-ph/9806292].

\bibitem{Bachas:1998kr}
C.~P.~Bachas,
JHEP {\bf 9811} (1998) 023
[hep-ph/9807415].

\bibitem{Antoniadis:1999ax}
I.~Antoniadis and C.~Bachas,
Phys.\ Lett.\ B {\bf 450} (1999) 83
[hep-th/9812093].

\bibitem{Arkani-Hamed:1999yp}
N.~Arkani-Hamed, S.~Dimopoulos and J.~March-Russell,
hep-th/9908146.

\bibitem{Krauss:1989zc}
L.~M.~Krauss and F.~Wilczek,
Phys.\ Rev.\ Lett.\  {\bf 62} (1989) 1221.

\bibitem{Alford:1990ch}
M.~G.~Alford, J.~March-Russell and F.~Wilczek,
Nucl.\ Phys.\ B {\bf 337} (1990) 695.

\bibitem{Preskill:1990bm}
J.~Preskill and L.~M.~Krauss,
Nucl.\ Phys.\ B {\bf 341} (1990) 50.

\bibitem{Alford:1991pt}
M.~G.~Alford, S.~Coleman and J.~March-Russell,
Nucl.\ Phys.\ B {\bf 351} (1991) 735.



\bibitem{Kamionkowski:1992mf}
M.~Kamionkowski and J.~March-Russell,
Phys.\ Lett.\ B {\bf 282} (1992) 137
[hep-th/9202003].

\bibitem{Israel:1966rt}
W.~Israel,
Nuovo Cim.\ B {\bf 44S10} (1966) 1.

\bibitem{Berezin:1987bc}
V.~A.~Berezin, V.~A.~Kuzmin and I.~I.~Tkachev,
Phys.\ Rev.\ D {\bf 36} (1987) 2919.

\bibitem{Freedman:2000gk}
D.~Z.~Freedman, S.~S.~Gubser, K.~Pilch and N.~P.~Warner,
JHEP{\bf 0007} (2000) 038
[hep-th/9906194].

\bibitem{Gogberashvili:2000iu}
M.~Gogberashvili,
Europhys.\ Lett.\ {\bf 49} (2000) 396
[hep-ph/9812365].

\bibitem{Gogberashvili:1999tb}
M.~Gogberashvili,
Mod.\ Phys.\ Lett.\ A {\bf 14} (1999) 2025
[hep-ph/9904383].

\bibitem{RS1}
L.~Randall and R.~Sundrum,
Phys.\ Rev.\ Lett.\ {\bf 83} (1999) 3370
[hep-ph/9905221].

\bibitem{RS2}
L.~Randall and R.~Sundrum,
Phys.\ Rev.\ Lett.\ {\bf 83} (1999) 4690
[hep-th/9906064].

\bibitem{Pilo}
L.~Pilo, R.~Rattazzi and A.~Zaffaroni,
JHEP{\bf 0007} (2000) 056
[hep-th/0004028].

\bibitem{Lisa}
C.~Csaki, M.~Graesser, L.~Randall and J.~Terning,
Phys.\ Rev.\ D {\bf 62} (2000) 045015
[hep-ph/9911406].

\bibitem{CGR}
C.~Charmousis, R.~Gregory and V.~A.~Rubakov,
Phys.\ Rev.\ D {\bf 62} (2000) 067505
[hep-th/9912160].

\bibitem{Goldberger:1999uk}
W.~D.~Goldberger and M.~B.~Wise,
Phys.\ Rev.\ Lett.\ {\bf 83} (1999) 4922
[hep-ph/9907447].

\bibitem{Luty:2000cz}
M.~A.~Luty and R.~Sundrum,
Phys.\ Rev.\ D {\bf 62} (2000) 035008
[hep-th/9910202].

\bibitem{Garriga:2000yh}
J.~Garriga and T.~Tanaka,
Phys.\ Rev.\ Lett.\ {\bf 84} (2000) 2778
[hep-th/9911055].

\bibitem{Giddings:2000mu}
S.~B.~Giddings, E.~Katz and L.~Randall,
JHEP {\bf 0003} (2000) 023
[hep-th/0002091].

\bibitem{Davoudiasl:2000jd}
H.~Davoudiasl, J.~L.~Hewett and T.~G.~Rizzo,
Phys.\ Rev.\ Lett.\ {\bf 84} (2000) 2080
[hep-ph/9909255].

\bibitem{Arkani-Hamed:2000ds}
N.~Arkani-Hamed, M.~Porrati and L.~Randall,
``Holography and phenomenology,''
hep-th/0012148.

\bibitem{Gherghetta:2000kr}
T.~Gherghetta and A.~Pomarol,
``A warped supersymmetric standard model,''
hep-ph/0012378.

\bibitem{Akhmedov:1999rc}
E.~T.~Akhmedov,
``Introduction to the AdS/CFT correspondence,''
hep-th/9911095, to appear in Uspekhi Fiz. Nauk.

\bibitem{Muck:2000bb}
W.~Muck, K.~S.~Viswanathan and I.~V.~Volovich,
Phys.\ Rev.\ D {\bf 62} (2000) 105019
[hep-th/0002132].

\bibitem{Gregory:2000rh}
R.~Gregory, V.~A.~Rubakov and S.~M.~Sibiryakov,
Class.\ Quant.\ Grav.\ {\bf 17} (2000) 4437
[hep-th/0003109].

\bibitem{Lykken:2000nb}
J.~Lykken and L.~Randall,
JHEP{\bf 0006} (2000) 014
[hep-th/9908076].

\bibitem{Hebecker:2001nv}
A.~Hebecker and J.~March-Russell,
``Randall-Sundrum II cosmology, AdS/CFT, and the bulk black hole,''
hep-ph/0103214.

\bibitem{DRT1}
S.~L.~Dubovsky, V.~A.~Rubakov and P.~G.~Tinyakov,
Phys.\ Rev.\ D {\bf 62} (2000) 105011
[hep-th/0006046].

\bibitem{Bajc}
B.~Bajc and G.~Gabadadze,
Phys.\ Lett.\ B {\bf 474} (2000) 282
[hep-th/9912232].

\bibitem{Maldacena:1998re}
J.~Maldacena,
Adv.\ Theor.\ Math.\ Phys.\ {\bf 2} (1998) 231
[hep-th/9711200].

\bibitem{Gubser:1998bc}
S.~S.~Gubser, I.~R.~Klebanov and A.~M.~Polyakov,
Phys.\ Lett.\ B {\bf 428} (1998) 105
[hep-th/9802109].

\bibitem{Witten:1998zw}
E.~Witten,
Adv.\ Theor.\ Math.\ Phys.\ {\bf 2} (1998) 505
[hep-th/9803131].

\bibitem{ads}
H.~Verlinde, Talk at ITP Santa Barbara conference ``New Dimensions
in Field Theory and String Theory'', 
http://www.itp.ucsb.edu/online/susy\_c99/verlinde ;  \\
E.~Witten, ibid., http://www.itp.ucsb.edu/online/susy\_c99/discussion .

\bibitem{Gubser:1999vj}
S.~S.~Gubser,
``AdS/CFT and gravity,''
hep-th/9912001.


\bibitem{Giddings:2000ay}
S.~B.~Giddings and E.~Katz,
hep-th/0009176.

\bibitem{Coleman:1977yb}
S.~Coleman and L.~Smarr,
Commun.\ Math.\ Phys.\ {\bf 56} (1977) 1.

\bibitem{Hawking:2000kj}
S.~W.~Hawking, T.~Hertog and H.~S.~Reall,
Phys.\ Rev.\ D {\bf 62} (2000) 043501
[hep-th/0003052].


\bibitem{Nojiri:2000gb}
S.~Nojiri and S.~D.~Odintsov,
Phys.\ Lett.\ B {\bf 484} (2000) 119
[hep-th/0004097].

\bibitem{Anchordoqui:2000du}
L.~Anchordoqui, C.~Nunez and K.~Olsen,
JHEP {\bf 0010} (2000) 050
[hep-th/0007064].

\bibitem{Arkani-Hamed:2000hk}
N.~Arkani-Hamed, S.~Dimopoulos, G.~Dvali and N.~Kaloper,
Phys.\ Rev.\ Lett.\ {\bf 84} (2000) 586
[hep-th/9907209].

\bibitem{Gherghetta:2000qi}
T.~Gherghetta and M.~Shaposhnikov,
Phys.\ Rev.\ Lett.\ {\bf 85} (2000) 240
[hep-th/0004014].

\bibitem{Cohen:1999ia}
A.~G.~Cohen and D.~B.~Kaplan,
Phys.\ Lett.\ B {\bf 470} (1999) 52
[hep-th/9910132].

\bibitem{Gregory:2000gv}
R.~Gregory,
Phys.\ Rev.\ Lett.\  {\bf 84} (2000) 2564
[hep-th/9911015].


\bibitem{Gherghetta:2000jf}
T.~Gherghetta, E.~Roessl and M.~Shaposhnikov,
Phys.\ Lett.\ B {\bf 491} (2000) 353
[hep-th/0006251].

\bibitem{Randjbar-Daemi:2000ft}
S.~Randjbar-Daemi and M.~Shaposhnikov,
Phys.\ Lett.\ B {\bf 491} (2000) 329
[hep-th/0008087].

\bibitem{DRT2}
S.~L.~Dubovsky, V.~A.~Rubakov and P.~G.~Tinyakov,
JHEP{\bf 0008} (2000) 041
[hep-ph/0007179].

\bibitem{Oda:2000zc}
I.~Oda,
Phys.\ Lett.\ B {\bf 496} (2000) 113
[hep-th/0006203].

\bibitem{Ponton:2001gi}
E.~Ponton and E.~Poppitz,
JHEP{\bf 0102} (2001) 042
[hep-th/0012033].

\bibitem{RubSh2}
V.~A.~Rubakov and M.~E.~Shaposhnikov,
Phys.\ Lett.\ B {\bf 125} (1983) 139.

\bibitem{Randjbar-Daemi:1986wg}
S.~Randjbar-Daemi and C.~Wetterich,
Phys.\ Lett.\ B {\bf 166} (1986) 65.

\bibitem{Arkani-Hamed:2000eg}
N.~Arkani-Hamed, S.~Dimopoulos, N.~Kaloper and R.~Sundrum,
Phys.\ Lett.\ B {\bf 480} (2000) 193
[hep-th/0001197].


\bibitem{Kachru:2000xs}
S.~Kachru, M.~Schulz and E.~Silverstein,
Phys.\ Rev.\ D {\bf 62} (2000) 085003
[hep-th/0002121].

\bibitem{Dolgov:1982gh}
A.~D.~Dolgov,
{\it  In *Cambridge 1982, Proceedings, The Very Early Universe*,
449-458}.

\bibitem{Peccei}
R.~D.~Peccei, J.~Sola and C.~Wetterich,
Phys.\ Lett.\ B {\bf 195} (1987) 183.


\bibitem{Barr}
S.~M.~Barr and D.~Hochberg,
Phys.\ Lett.\ B {\bf 211} (1988) 49.

\bibitem{deAlwis}
S.~P.~de Alwis,
Nucl.\ Phys.\ B {\bf 597} (2001) 263
[hep-th/0002174]; \\
S.~P.~de Alwis, A.~T.~Flournoy and N.~Irges,
JHEP{\bf 0101} (2001) 027
[hep-th/0004125].

\bibitem{Forste:2000ps}
S.~Forste, Z.~Lalak, S.~Lavignac and H.~P.~Nilles,
Phys.\ Lett.\ B {\bf 481} (2000) 360
[hep-th/0002164].

\bibitem{Csaki:2000wz}
C.~Csaki, J.~Erlich, C.~Grojean and T.~Hollowood,
Nucl.\ Phys.\ B {\bf 584} (2000) 359
[hep-th/0004133].

\bibitem{Rubakov:2000aq}
V.~A.~Rubakov,
Phys.\ Rev.\ D {\bf 61} (2000) 061501
[hep-ph/9911305].

\bibitem{Kogan:2000wc}
I.~I.~Kogan, S.~Mouslopoulos, A.~Papazoglou, G.~G.~Ross and J.~Santiago,
Nucl.\ Phys.\ B {\bf 584} (2000) 313
[hep-ph/9912552].

\bibitem{GRS}
R.~Gregory, V.~A.~Rubakov and S.~M.~Sibiryakov,
Phys.\ Rev.\ Lett.\ {\bf 84} (2000) 5928
[hep-th/0002072].


\bibitem{Csaki:2000pp}
C.~Csaki, J.~Erlich and T.~J.~Hollowood,
Phys.\ Rev.\ Lett.\ {\bf 84} (2000) 5932
[hep-th/0002161].

\bibitem{Dvali:2000rv}
G.~Dvali, G.~Gabadadze and M.~Porrati,
Phys.\ Lett.\ B {\bf 484} (2000) 112
[hep-th/0002190].

\bibitem{Gregory:2000iu}
R.~Gregory, V.~A.~Rubakov and S.~M.~Sibiryakov,
Phys.\ Lett.\ B {\bf 489} (2000) 203
[hep-th/0003045].



\bibitem{Csaki:2001cx}
C.~Csaki, J.~Erlich, T.~J.~Hollowood and J.~Terning,
Phys.\ Rev.\ D {\bf 63} (2001) 065019
[hep-th/0003076].

\bibitem{Witten:2000zk}
E.~Witten,
``The cosmological constant from the viewpoint of string theory,''
hep-ph/0002297.

\bibitem{Kogan:2000cv}
I.~I.~Kogan and G.~G.~Ross,
Phys.\ Lett.\ B {\bf 485} (2000) 255
[hep-th/0003074].

\bibitem{Dvali:2000km}
G.~Dvali, G.~Gabadadze and M.~Porrati,
Phys.\ Lett.\ B {\bf 484} (2000) 129
[hep-th/0003054].


\bibitem{vanDam:1970vg}
H.~van Dam and M.~Veltman,
Nucl.\ Phys.\ B {\bf 22} (1970) 397.

\bibitem{Zakharov}
V.~I.~Zakharov,
JETP Lett.\  {\bf 12}, 312 (1970).

\bibitem{Kogan:2001uy}
I.~I.~Kogan, S.~Mouslopoulos and A.~Papazoglou,
Phys.\ Lett.\ B {\bf 503} (2001) 173
[hep-th/0011138].

\bibitem{Porrati:2001cp}
M.~Porrati,
Phys.\ Lett.\ B {\bf 498} (2001) 92
[hep-th/0011152].

\bibitem{Dilkes:2001av}
F.~A.~Dilkes, M.~J.~Duff, J.~T.~Liu and H.~Sati,
``Quantum M --> 0 discontinuity for massive gravity with a Lambda term,''
hep-th/0102093.

\bibitem{Kogan:2001vb}
I.~I.~Kogan, S.~Mouslopoulos and A.~Papazoglou,
Phys.\ Lett.\ B {\bf 501} (2001) 140
[hep-th/0011141].


\bibitem{Karch:2000ct}
A.~Karch and L.~Randall,
``Locally localized gravity,''
hep-th/0011156.

\bibitem{viss}
M.~Visser,
Phys.\ Lett.\ B {\bf 159} (1985) 22.


\bibitem{Bowcock:2000cq}
P.~Bowcock, C.~Charmousis and R.~Gregory,
Class.\ Quant.\ Grav.\  {\bf 17} (2000) 4745
[hep-th/0007177].

\bibitem{Chung:2000ji}
D.~J.~Chung, E.~W.~Kolb and A.~Riotto,
``Extra dimensions present a new flatness problem,''
hep-ph/0008126.

\bibitem{Csaki:2000dm}
C.~Csaki, J.~Erlich and C.~Grojean,
``Gravitational Lorentz violations and adjustment of the cosmological  constant in asymmetrically warped spacetimes,''
hep-th/0012143.

\bibitem{Dubovsky:2001fj}
S.~L.~Dubovsky,
``Tunneling into extra dimension and high-energy 
violation of Lorentz  invariance,''
hep-th/0103205.


\bibitem{Coleman:1999ti}
S.~Coleman and S.~L.~Glashow,
Phys.\ Rev.\ D {\bf 59} (1999) 116008
[hep-ph/9812418].

\bibitem{Csaki:1999jh}
C.~Csaki, M.~Graesser, C.~Kolda and J.~Terning,
Phys.\ Lett.\ B {\bf 462} (1999) 34
[hep-ph/9906513].

\bibitem{Cline:1999ts}
J.~M.~Cline, C.~Grojean and G.~Servant,
Phys.\ Rev.\ Lett.\ {\bf 83} (1999) 4245
[hep-ph/9906523].

\bibitem{Csaki:2000mp}
C.~Csaki, M.~Graesser, L.~Randall and J.~Terning,
Phys.\ Rev.\ D {\bf 62} (2000) 045015
[hep-ph/9911406].

\bibitem{Binetruy:2000ut}
P.~Binetruy, C.~Deffayet and D.~Langlois,
Nucl.\ Phys.\ B {\bf 565} (2000) 269
[hep-th/9905012].

\bibitem{Chung:2000zs}
D.~J.~Chung and K.~Freese,
Phys.\ Rev.\ D {\bf 61} (2000) 023511
[hep-ph/9906542].

\bibitem{Gorsky:2000rz}
A.~Gorsky and K.~Selivanov,
Phys.\ Lett.\ B {\bf 485} (2000) 271
[hep-th/0005066]; 
``Brane tunneling and the brane world scenario,''
hep-th/0009207.

\bibitem{Brown:1988kg}
J.~D.~Brown and C.~Teitelboim,
Nucl.\ Phys.\ B {\bf 297} (1988) 787.

\end{thebibliography}
\end{document}